\documentclass{aa}  

\usepackage{graphicx}
\usepackage{txfonts}
\usepackage{multirow}
\usepackage{natbib}
 \usepackage[colorlinks=true,linkcolor=blue,citecolor=blue]{hyperref} 
\bibpunct{(}{)}{;}{a}{}{,} 

\makeatletter
\renewcommand*\aa@pageof{, page \thepage{} of \pageref*{LastPage}}
\makeatother

\usepackage[table]{xcolor}
\definecolor{sronred}{RGB}{220, 5, 12}
\definecolor{sronblue}{RGB}{25, 101, 176}

\begin{document}

\title{Influence of star-forming galaxy selection on the galaxy main sequence}
\subtitle{}

\author{W.~J.~Pearson\inst{\ref{inst:NCBJ}}
      \and F.~Pistis\inst{\ref{inst:NCBJ}}
      \and M.~Figueira\inst{\ref{inst:NCBJ}, \ref{inst:NCU}}
      \and K.~Ma\l ek\inst{\ref{inst:NCBJ}, \ref{inst:LAM}}
      \and{T. Moutard}\inst{\ref{inst:LAM}}
      \and D.~Vergani\inst{\ref{inst:INAF}}
      \and A.~Pollo\inst{\ref{inst:NCBJ}, \ref{inst:Jag}}
      }

\institute{National Centre for Nuclear Research, Pasteura 7, 02-093 Warszawa, Poland\label{inst:NCBJ}\\\email{william.pearson@ncbj.gov.pl}
	\and
        Institute of Astronomy, Faculty of Physics, Astronomy and Informatics, Nicolaus Copernicus University, Grudzi\k{a}dzka 5, 87-100 Toruń, Poland\label{inst:NCU}
        \and
         Aix Marseille Univ. CNRS, CNES, LAM, Marseille, France\label{inst:LAM}
        \and
        INAF – Osservatorio di Astrofisica e Scienza dello Spazio, Via P.Gobetti 93/3, 40129 Bologna, Italy\label{inst:INAF}
        \and
Astronomical Observatory of the Jagiellonian University, ul. Orla 171, 30-244 Kraków, Poland\label{inst:Jag}
}

\date{Received DD Month YYYY; accepted DD Month YYYY}

 
\abstract
{}
{This work aims to determine how the galaxy main sequence (MS) changes using seven different commonly used methods to select the star-forming galaxies within VIPERS data over $0.5 \leq z < 1.2$. The form and redshift evolution of the MS will then be compared between selection methods.}
{The star-forming galaxies were selected using widely known methods: a specific star-formation rate (sSFR), Baldwin, Phillips and Terlevich (BPT) diagram, 4000\AA\ spectral break (D4000) cut and four colour-colour cuts:  NUVrJ, NUVrK, u-r, and UVJ. The main sequences were then fitted for each of the seven selection methods using a Markov chain Monte Carlo forward modelling routine, fitting both a linear main sequence and a MS with a high-mass turn-over to the star-forming galaxies. This was done in four redshift bins of $0.50 \leq z < 0.62$, $0.62 \leq z < 0.72$, $0.72 \leq z < 0.85$, and $0.85 \leq z < 1.20$.}
{The slopes of all star-forming samples were found to either remain constant or increase with redshift, and the scatters were approximately constant. There is no clear redshift dependency of the presence of a high-mass turn-over for the majority of samples, with the NUVrJ and NUVrK being the only samples with turn-overs only at low redshift. No samples have turn-overs at all redshifts. Star-forming galaxies selected with sSFR and u-r are the only samples to have no high-mass turn-over in all redshift bins. The normalisation of the MS increases with redshift, as expected. The scatter around the MS is lower than the $\approx$0.3~dex typically seen in MS studies for all seven samples.}
{The lack, or presence, of a high-mass turn-over is at least partially a result of the method used to select star-forming galaxies. However, whether a turn-over should be present or not is unclear.}


\maketitle

\section{Introduction}
The galaxy main sequence (MS) is an observed tight correlation between the star-formation rate (SFR) and M$_{\star}$ (M$_{\star}$) of star-forming galaxies \citep{2004MNRAS.351.1151B, 2007ApJ...660L..43N, 2007A&A...468...33E}. The scatter of this relation is found to be approximately 0.2-0.3~dex, independent of M$_{\star}$, and remarkably consistent across the majority of the history of the universe \citep[e.g.][]{2012ApJ...754L..29W, 2014ApJS..214...15S, 2016ApJ...820L...1K, 2016ApJ...817..118T, 2018A&A...615A.146P}. This scatter is believed to be a result of fluctuations of in-falling material onto galaxies and periods of bursty star-formation \citep[e.g.][]{2014ApJ...785L..36A, 2016MNRAS.457.2790T, 2017MNRAS.464.2766M}. The consistency of the scatter is seen to be a result of SF being dominated by similar, slow processes of gradual M$_{\star}$ growth in all galaxies at all cosmic times \citep{2015ApJ...801...80L}.

The slope of the MS, or the low mass MS where a turn-over is seen, is typically found to be in the range 0.4 to 1.0 \citep{2014ApJ...795..104W, 2015A&A...575A..74S, 2016ApJ...817..118T, 2018A&A...615A.146P, 2023MNRAS.519.1526P}. The slope has been seen to reduce towards higher redshift in a small number of studies \citep[e.g.][]{2020MNRAS.499..948R} while predominantly being seen to increase with redshift in others \citep[e.g.][]{2014ApJS..214...15S, 2018A&A...615A.146P}. Further works argue that the slope of the MS should be unity at all redshifts \citep{2017ApJ...834...39P, 2023MNRAS.519.1526P}. Slopes of less than unity are considered to be a result of the relative reduction in size of the cold gas reservoir as M$_{\star}$ increases along with star-formation quenching \citep{2017ApJ...834...39P}.

Current literature tends to agree on the evolution of the normalisation of the MS. Numerous studies have shown that the normalisation increases with redshift \citep[e.g.][]{2014ApJS..214...15S, 2015A&A...575A..74S, 2016ApJ...817..118T, 2018A&A...615A.146P, 2023MNRAS.519.1526P}. This decrease in SFR as the universe ages may be a result of the availability of cool gas reducing as the redshift decreases \citep{2010Natur.463..781T, 2011MNRAS.417.1510D, 2015ApJ...800...20G, 2016ApJ...820...83S, 2021ApJ...921...40K}. This, coupled with the SFR per dust mass either being constant or increasing with redshift \citep{2010Natur.463..781T, 2016ApJ...820...83S} implies an expected increase in SFR with redshift, and hence an increase in the normalisation of the MS with redshift.

The exact form of the MS is highly contested in literature. A number of studies have found that the MS is a simple linear power law of the form $\log(\mathrm{SFR}) \propto \log(\mathrm{M_{\star}})$ at all M$_{\star}$ \citep[e.g.][]{2014ApJS..214...15S, 2018A&A...615A.146P}. Meanwhile, other studies have found the MS to have a high-mass turn-over, with the MS becoming less steep at higher masses \citep[e.g.][]{2012ApJ...754L..29W, 2015ApJ...801...80L, 2016ApJ...817..118T, 2019MNRAS.483.3213P, 2023MNRAS.519.1526P}. This apparent discrepancy has been attributed to how the star-forming galaxies are selected. \citet{2015MNRAS.453.2540J} showed that with a more aggressive cut, that is one that has stricter criteria for a star-forming galaxy, the MS shows less evidence for a high-mass turn-over. Similarly, a less aggressive cut shows strong evidence for a turn-over. The turn-over has also been seen to be a result of the SFR tracer used, as well as the tracer of the MS itself \citep[mean, mode or median;][]{2019MNRAS.483.3213P}.

The selection of star-forming galaxies can be done in a number of different ways. A for example, simple cut in specific star-formation rate (sSFR) \citep[e.g.][]{2018ApJ...859...11S}. Alternatively, as star-forming galaxies are typically bluer in colour, a colour or colour-colour cut can be performed. These cuts, such as u-r \citep[e.g.][]{2015MNRAS.453.2540J} or UVJ \citep[e.g.][]{2011ApJ...735...86W, 2014ApJ...795..104W}, split the red galaxies, assumed to be quiescent, from the blue galaxies, assumed to be star-forming. These colour or colour-colour cuts are often determined semi-visually, by populating the parameters space and observing a clustering of the red and blue galaxies. A line is then drawn between these two clusters to separate the quiescent galaxies from the star-forming galaxies. There are a number of such cuts using different colours but all functioning using the same underlying logic. Separation of star-forming and quiescent galaxies can also be performed using spectroscopic observations, such as with the \citet[][BPT]{1981PASP...93....5B} diagram, which requires emission lines measurements, or using the strongest discontinuity in the optical spectrum of a galaxy - the 4000~\AA{} break \citep[D4000,][]{1999ApJ...527...54B, 2005MNRAS.362...41G}

In this paper, we study the form of the MS in the VIMOS Public Extragalactic Redshift Survey\footnote{http://vipers.inaf.it/} \citep[VIPERS,][]{garilli14, Guzzo2014,scodeggio18} using different methods to select the star-forming galaxies, both photometric and spectroscopic. The resulting MSs will be compared to understand how selection influences the shape and apparent evolution of the MS. This may help to understand why some studies find a high-mass turn-over in the MS while others do not.

The paper is structured as follows. Section \ref{sec:data} describes the data used along with the star-forming galaxy selection methods, Sect. \ref{sec:methods} explains the methodology used to fit the MS, Sect. \ref{sec:result} presents the results, and Sect. \ref{sec:discuss} provides discussion. We summarise and conclude in Sect. \ref{sec:conclusion}.
 
\section{Data}\label{sec:data}
\subsection{VIPERS}
VIPERS is a completed ESO Large Program, which aimed to investigate the spatial distribution of galaxies over the $z$\textasciitilde1 Universe \citep{Guzzo2014,scodeggio18}. 
VIPERS was performed by the Visible Multi-Object Spectrograph \citep[VIMOS,][]{lefevre03} at moderate resolution (R = 220), using the LR red grism and providing a wavelength coverage of 5500–9500~\AA.
The galaxy target sample was selected from optical photometric catalogues of the Canada-France-Hawaii Telescope Legacy Survey Wide (CFHTLS-Wide), and covered $\sim$23.5~$\deg^2$ on the sky. 
The survey was divided into two areas within the W1 and W4 CFHTLS fields. 
A simple and robust colour preselection 
\begin{equation}
(r-i) > 0.5(u-g)\mbox{ OR }(r-i) > 0.7,   
\end{equation}
was  applied to efficiently remove galaxies at $z < 0.5$. 
Moreover, criterion of $i_{AB} < 22.5$~mag was used to select target sample. 
Finally VIPERS provided spectroscopic redshifts, spectra and full photometrically-selected parent catalogue for 86\,082 (+ 530 secondary objects)  galaxies.

VIPERS obtained a large volume of 5$\times$10$^7$ $\rm h^{-3}Mpc^{3}$ and an average target sampling rate of $>45$\%. This combination of sampling and depth is uncommon for intermediate redshift surveys at $z > 0.5$ and places VIPERS as one of a very small number of counterparts of local spectroscopic surveys like SDSS. Therefore VIPERS measurements are perfect to study the evolution of the MS above the local Universe traced by the SDSS. 

A~detailed description of the survey design and final results can be found in~\cite{Guzzo2014}, and \cite{scodeggio18}, respectively.
The data reduction pipeline and redshift quality system are described by \cite{garilli14}. In this work, we use galaxies with spectroscopic redshift quality flag between 3.0 and 4.5, which have a redshift confidence level greater than 95\% \citep{garilli14, scodeggio18}. The same selection was made in \citet{figueira2022sfr} to constrain the sets of the SFR calibrators for star forming galaxies and in \citet{2017A&A...597A.107S} to discuss the star formation history of passive red galaxies. Applying these cuts gives us a sample of 29\,958 galaxies that we use for further analysis.

\subsection{Stellar masses and star-formation rates}
The stellar masses and SFRs used in this work are derived through spectral energy distribution (SED) modelling using Code Investigating GALaxy Emission \citep[CIGALE,][]{2009A&A...507.1793N, 2019A&A...622A.103B} version \texttt{2022.1}, fitting to broad band photometry from the far-ultraviolet to the far-infrared, where available. We use a \citet{2003MNRAS.344.1000B} stellar population, \citet{2003PASP..115..763C} initial mass function, \citet{2000ApJ...539..718C} dust attenuation, \citet{2014ApJ...780..172D} update to the \citet{2007ApJ...657..810D} dust emission, and \citet{2006MNRAS.366..767F} active galactic nuclei (AGN) emission. For the star-formation history (SFH) we use a non-parametric model that does not assume a specific shape of the SFH \citep{2023A&A...672A.191C}. A non-parametric model is used to minimise the degeneracy that there is between M$_{\star}$ and SFR in SED fitting with parametric SFH models, which can result in bands being formed in the SFR-M$_{\star}$ plane which may influence our results. The parameters used for fitting can be found in Appendix \ref{app:cigale}.

\subsection{Emission Lines}
We perform the analysis of the observed VIPERS spectra (Pistis et al. 2023, in prep.) via the penalized pixel fitting code \citep[pPXF][]{cappellari2004fit, cappellari2017improving, cappellari2022full}. 
pPXF performs the fit of stellar templates, based on the library MILES \citep{vazdekis2010miles}, and the fit of the gas templates, with the emission lines present in the observed spectral region, separately.
The gas component is then fitted with a single Gaussian for each emission line giving the integrated fluxes and their errors as a result.
For better estimation of the errors, the errors given by pPXF are multiplied by the $\chi^2_\mathrm{red}$ of the fit below the line.

To estimate the equivalent widths (EWs) and their error, we build a spectrum with a continuum normalized by dividing the total fit of the spectrum by the fit of the stellar component from pPXF.
The resulting normalized spectrum is analysed with specutils, an astropy package for spectroscopy \citep{astropy2013, astropy2018, price2018astropy} in a range of $\pm 1.06$ full width half maximum (FWHM), which is equivalent to $5 \sigma$ of the Gaussian fit \citep{vietri2022agn}, around the centroid of the emission line.

The quality of the emission line measurements is given by a quality flag \citep{garilli2010ez, figueira2022sfr, pistis2022bias} in the form of $xyzt$.
The $x$ value flags lines where the centroid is less that 7~\AA\, the $y$ value flags lines where the FWHM is between 7 and 22~\AA\, the $z$ value flags lines where the difference between the peak and the fit of the line is more than 30\%, and the $t$ flags galaxies with signal-to-noise (S/N) ratio of the EW is at least 3.5, the S/N of the flux is at least 8, or the S/N of the flux is at least 7.
In this work for the BPT sample we require the galaxy to have at least one flag set to zero for all three required lines: H$\beta$, [\ion{O}{iii}] $\lambda$5007, and [\ion{O}{ii}] $\lambda\lambda$3727.

\subsection{Star-forming galaxy selection}
In this work we perform seven selections for star-forming galaxies: photometrically we perform a cut in sSFR, a NUVrJ colour cut, a NUVrK colour cut, a u-r colour cut, and a UVJ colour cut, and spectroscopically we perform BPT selection and a D4000 cut. For all the selections, we enforce that galaxies must have SFR and M$_{\star}$ values as well as the required bands, lines, or rest frame colours for that selection.

The data are also binned by redshift. To determine the edges of the redshift bins, we split the sSFR sample into four bins containing approximately an equal number of galaxies. This results in bins with edges at $z=0.50$, $z=0.62$, $z=0.72$, $z=0.85$, and $z=1.20$. The differing requirements, described below, will result in different sample sizes for the different selection methods. The sample sizes for each selection method in each redshift bin can be found in Tables \ref{tab:3param} and \ref{tab:4param}.

We note, however, that any star-forming galaxy selection made will be imperfect. There can be contamination from quiescent galaxies as well as truly star-forming galaxies that are not classified as such, which will vary between the methods used to identify star-forming galaxies. The star-forming galaxy samples can also include transient/E+A galaxies, which are known to contribute up to 8\% of quiescent galaxy samples \citep{2010A&A...509A..42V} at $0.48 < z < 1.20$ and likely similarly contaminate star-forming galaxy samples.

\subsubsection{sSFR}
\begin{figure}
    \resizebox{\hsize}{!}{\includegraphics{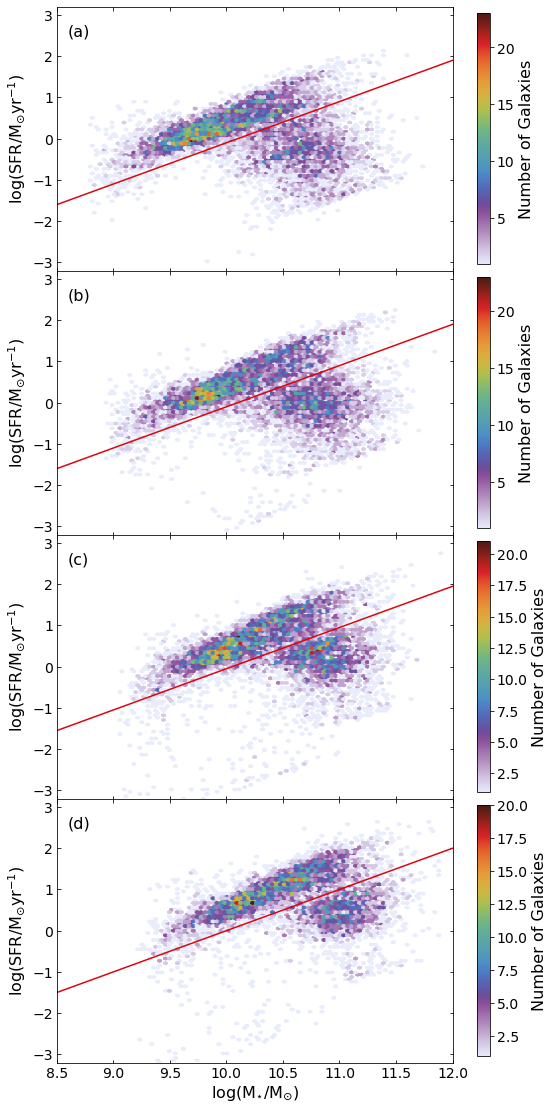}}
    \caption{Density plot of SFR vs M$_{\star}$ for galaxies with (a) $0.50 \leq z < 0.62$, (b) $0.62 \leq z < 0.72$, (c) $0.72 \leq z < 0.85$, and (d) $0.85 \leq z < 1.20$ from low density (light purple) to high density (dark red). The sSFR cuts are shown as a red lines.}
    \label{fig:sSFR-cut}
\end{figure}

For the cut in sSFR, we visually inspect the SFR-M$_{\star}$ plane, as can be seen in Fig. \ref{fig:sSFR-cut}, and define star-forming galaxies as:
\begin{equation}\label{eqn:sSFR}
    \begin{aligned}
        \log(\mathrm{sSFR/yr}) > -10.10 && 0.50 \leq z < 0.62\\
        \log(\mathrm{sSFR/yr}) > -10.10 && 0.62 \leq z < 0.72\\
        \log(\mathrm{sSFR/yr}) > -10.05 && 0.72 \leq z < 0.85\\
        \log(\mathrm{sSFR/yr}) > -10.00 && 0.85 \leq z < 1.20\\
    \end{aligned}
\end{equation}
All 29\,958 galaxies in our sample have the required SFR and M$_{\star}$, of which 18\,060 were selected as star-forming.

\subsubsection{NUVrJ}
For NUVrJ we follow \citet{2013A&A...556A..55I} to select star-forming galaxies. Thus, we select star-forming galaxies to have:
\begin{equation} \label{eqn:NUVrJ}
\begin{split}
    NUV - r &< 3(r - J) + 1 \mathrm{~or,}\\
    NUV - r &< 3.1.
\end{split}
\end{equation}
All 29\,958 galaxies of our sample have the required NUV, r, and J observations, of which 22\,868 were selected as star-forming. The selection plot can be found in Appendix \ref{app:sf-plots}.

It is possible that the NUVrJ sample can become contaminated with green valley galaxies as there is no clear delineation between star-forming and green valley galaxies in the NUVrJ colour-colour space \citep{2010ApJ...709..644I, 2020MNRAS.495.4237M}.

\subsubsection{NUVrK}
For NUVrK we select the star-forming galaxies as has been done by \citet{Davidzon2016}: 
\begin{equation} \label{eqn:NUVrK}
\begin{split}
    NUV - r &< 1.37(r - K) + 2.6 \mathrm{~or,}\\
    NUV - r &< 3.15 \mathrm{~or,}\\
    r - K &> 1.3.
\end{split}
\end{equation}
All 29\,958 galaxies of our sample have the required NUV, r, and K observations, of which 22\,934 were selected as star-forming. The selection plot can be found in Appendix \ref{app:sf-plots}.

While similar to NUVrJ, the r-K colour becomes redder with cosmic time, and is also sensitive to the inclination of a galaxy. Together with NUV-r it is therefore a good tracer of the sSFR that can be better than NUVrJ at separating active galaxies \citep{2018A&A...617A..70S, 2020MNRAS.495.4237M}. It is also easy for the NUVrK sample to become contaminated with green valley galaxies as there is no clear delineation between star-forming and green valley galaxies \citep{2020MNRAS.495.4237M}.

\subsubsection{u-r}
\begin{figure}
    \resizebox{0.95\hsize}{!}{\includegraphics{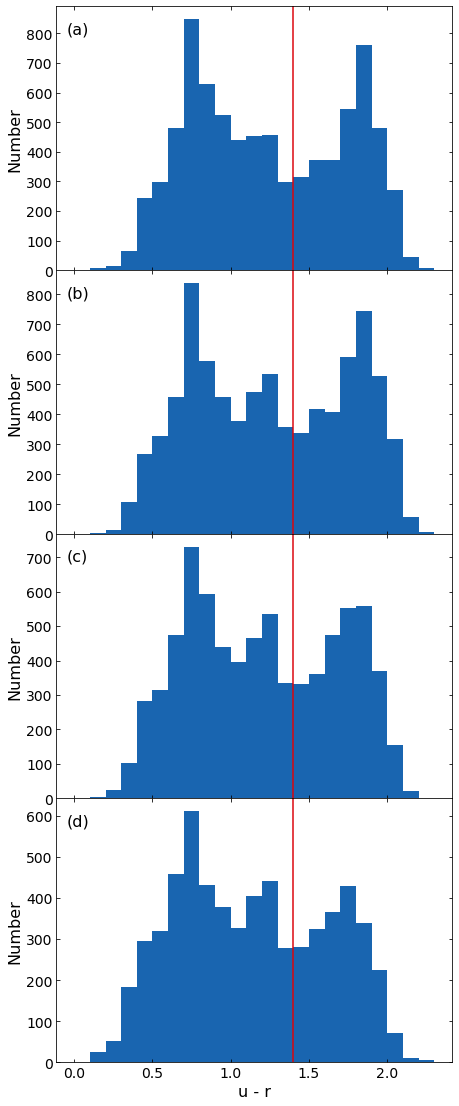}}
    \caption{Histogram of rest-frame u-r colour for (a) $0.50 \leq z < 0.62$, (b) $0.62 \leq z < 0.72$, (c) $0.72 \leq z < 0.85$, and (d) $0.85 \leq z < 1.20$. Red lines show the split between star-forming and quiescent galaxies.}
    \label{fig:ur-cut}
\end{figure}

We define the selection in u-r after visual examination of the distribution of the rest-frame u-r colour, which we estimate with CIAGLE, as shown in Fig. \ref{fig:ur-cut} \citep{2015MNRAS.453.2540J}. We select star-forming galaxies as those with $u-r < 1.4$ at all redshifts. All 29\,958 galaxies of our sample have the required rest-frame u-r colour, of which 18\,470 were selected as star-forming.

The cut we apply is a bluer cut than that applied by \citet{2015MNRAS.453.2540J}. As a result, we may be applying a stricter definition of star-forming galaxies and hence be removing more quiescent galaxies. The u-r colour is also sensitive to dust reddening and, as a result, may remove more dusty star-forming galaxies, which typically have a higher M$_{\star}$ \citep[e.g.][]{2020A&A...644A.144D}.

\subsubsection{UVJ}
For our final star-forming galaxy selection, we use the rest-frame UVJ colour cut of \citet{2011ApJ...735...86W} at low redshift:
\begin{equation}\label{eqn:UVJ}
    \begin{aligned}
        (U - V) &< 0.88 \times (V - J) + 0.59 \mathrm{~or,}\\
        (U - V) &< 1.30 \mathrm{~or,}\\
        (V - J) &> 1.60,
    \end{aligned}
\end{equation}
where the U-V and V-J rest-frame colours are estimated with CIGALE. All 29\,958 galaxies of our sample have the required rest-frame U-V and V-J colours, of which 20\,371 were selected as star-forming. The selection plot can be found in Appendix \ref{app:sf-plots}.

The UVJ selection method has been shown to separate quiescent and green-valley galaxies from star-forming galaxies \citep{2020MNRAS.495.4237M}. As a result, the UVJ selected star-forming sample should be a relatively pure sample.

\subsubsection{BPT}
The BPT sample is spectroscopically selected and, as such, contains far fewer galaxies than the photometrically selected samples previously discussed. It should, however, be a much purer star-forming sample than other methods. For BPT selected star-forming galaxies, we follow \citet{2010A&A...509A..53L} ``blue BPT''. The blue BPT selects galaxies using H$\beta$, [\ion{O}{iii}] $\lambda$5007, and [\ion{O}{ii}] $\lambda\lambda$3727 emission lines. We require that all emission lines have a flux signal-to-noise ratio greater than 3.
We define star-forming galaxies as galaxies with:
\begin{equation} \label{eqn:BPT}
\begin{split}
    \log([\mathrm{\ion{O}{iii}}]/\mathrm{H}\beta) &< \frac{0.11}{\log([\mathrm{\ion{O}{ii}}]/\mathrm{H}\beta) - 0.92} + 0.85,\\
    \log([\mathrm{\ion{O}{iii}}]/\mathrm{H}\beta) &< 0.3.
\end{split}
\end{equation}
Only 6310 galaxies of our sample have the required lines at the required SNR, of which 4560 were selected as star-forming. The selection plot can be found in Appendix \ref{app:sf-plots}.

\subsubsection{D4000}
For the final selection, cutting in D4000, we follow \citet{2017A&A...605A...4H} and define star-forming galaxies as:
\begin{equation}\label{eqn:D4000}
    \begin{aligned}
    &D4000 < 1.55,\\
    &\log(\mathrm{M}_{\star}/\mathrm{M}_{\odot}) < 11.0.
	\end{aligned}
\end{equation}
29\,955 galaxies of our sample have the required D4000 measurement, of which 22\,010 were selected as star-forming. The selection plot can be found in Appendix \ref{app:sf-plots}.

The limiting value of 1.55 for D4000 was tested  based on the sample of  VIPERS galaxies by \citet{2017A&A...605A...4H}  but also the same limit was used for local Universe SDSS galaxies by \citet{2003MNRAS.341...33K}. The D4000 index is much easier to measure than the selection of emission lines, as it is the manifestation of the strongest discontinuity in the continuum. The wavelength range of VIPERS (5500-9500\AA) makes the calculation of this estimator of if a galaxy is star-forming, or not, rather straightforward. Is it the reason why the sample selected based on the D4000 strength is much larger than the one based on the blue BPT diagram. The D4000 selection can also contain a large contamination of more passive galaxies in the star-forming sample, especially at higher M$_{\star}$ \citep{2008A&A...487...89V}. D4000 is also one of the measurements with lowest associated error.

\subsection{Sample differences}
As each star-forming galaxy sample is selected differently, it is informative to see the overlap between the different samples. In Table \ref{tab:overlap} we present the fraction of the sample selected by the column header that is contained within the sample selected by the header of the row. Due to the requirement that the BPT sample has signal-to-noise ratios greater than 3 for the four emission lines, H$\beta$, [\ion{O}{iii}] $\lambda$5007, and [\ion{O}{ii}] $\lambda$3727, this sample contains a small fraction of the other samples, at most 44\%. This limitation in sample size is also due to the observations of the required lines being limited to $0.5 < z < 0.9$. As a result, the results with the BPT selection should be treated with caution even though it should contain the most obvious sample of star-forming galaxies as all of them are characterized by a very strong set of emission lines. This strict requirement on emission lines reduces the completeness of the sample.

\begin{table*}
    \centering
    \caption{Fraction of the star-forming galaxy sample selected with the column header that is present in the star-forming galaxy sample selected with the row header.}
    \label{tab:overlap}
    \begin{tabular}{c|ccccccc}
        \hline
        \hline
         & sSFR & NUVrJ & NUVrK & u-r & UVJ & BPT & D4000 \\
        \hline
        sSFR & - & 0.78 & 0.78 & 0.89 & 0.86 & 0.78 & 0.83 \\
        NUVrJ & \cellcolor{sronblue!50}0.99 & - & \cellcolor{sronblue!50}0.99 & \cellcolor{sronblue!50}0.99 & \cellcolor{sronblue!50}0.98 & \cellcolor{sronblue!50}0.94 & \cellcolor{sronblue!50}0.98 \\
        NUVrK & \cellcolor{sronblue!50}0.99 & \cellcolor{sronblue!50}0.99 & - & \cellcolor{sronblue!50}0.99 & \cellcolor{sronblue!50}0.98 & \cellcolor{sronblue!50}0.94 & \cellcolor{sronblue!50}0.98 \\
        u-r & \cellcolor{sronblue!50}0.91 & 0.80 & 0.80 & - & 0.88 & 0.82 & 0.88 \\
        UVJ & \cellcolor{sronblue!50}0.97 & 0.87 & 0.87 & \cellcolor{sronblue!50}0.97 & - & 0.87 & \cellcolor{sronblue!50}0.92 \\
        BPT & \cellcolor{sronred!25}0.42 & \cellcolor{sronred!25}0.40 & \cellcolor{sronred!25}0.40 & \cellcolor{sronred!25}0.43 & \cellcolor{sronred!25}0.42 & - & \cellcolor{sronred!25}0.44 \\
        D4000 & 0.89 & 0.83 & 0.83 & \cellcolor{sronblue!50}0.92 & 0.87 & 0.87 & - \\
        \hline
    \end{tabular}
    \tablefoot{If more than 90\% of the column sample is present in the row sample, the cell is shaded dark blue while if less that 50\% of the column sample is present in the row sample, the cell is shaded light red.}
\end{table*}

\begin{figure}
	\resizebox{\hsize}{!}{\includegraphics{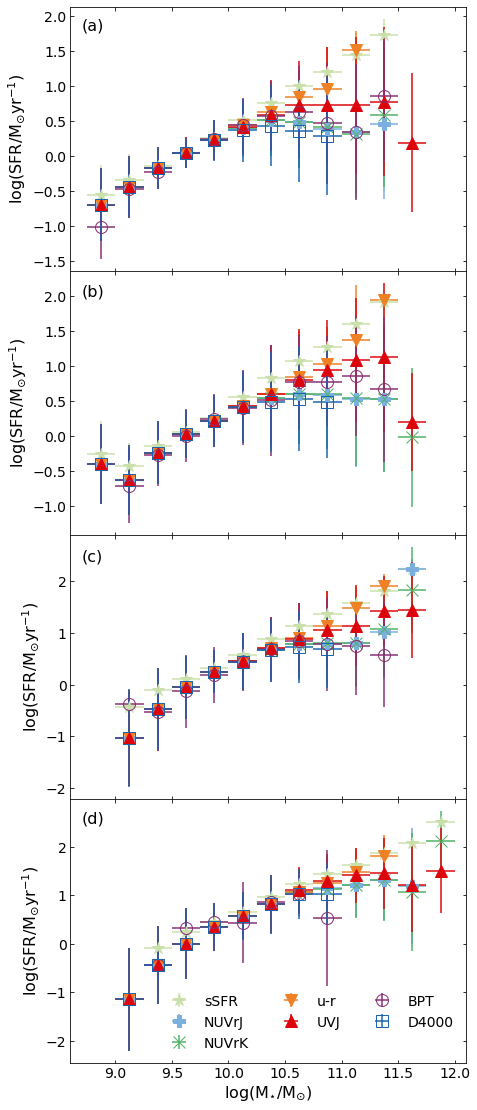}}
	\caption{The mean SFR and standard deviation (vertical error bars) of the sSFR (light green stars), NUVrJ (light blue pluses),  NVUrK (dark green crosses), u-r (orange, downward pointing triangles), UVJ (red, upward pointing triangles), BPT (empty purple circles), and D4000 (empty dark blue squares) selected star-forming galaxies at (a) $0.50 \leq z < 0.62$, (b) $0.62 \leq z < 0.72$, (c) $0.72 \leq z < 0.85$, and (d) $0.85 \leq z < 1.20$. Horizontal error bars indicate the width of the M$_{\star}$ bin.}
	\label{fig:binned-SFR}
\end{figure}

It is also informative to understand where the samples differ in the SFR-M$_{\star}$ plane. We calculate the mean and standard deviations of the SFR in mass bins of width 0.25~dex (the same M$_{\star}$ bins used in the forward modelling described in Sect. \ref{sec:forward-modelling}) and present them in Fig. \ref{fig:binned-SFR}. As can be seen, the samples are similar at lower M$_{\star}$ (log(M$_{\star}$/M$_{\odot}$ $\lessapprox 10.5$ depending on redshift). At higher mass, sSFR and u-r are more restrictive in selection while NUVrJ, NUVrK, BPT, and D4000 are less restrictive. The UVJ selection lies between these two extremes.

\section{Methods}\label{sec:methods}
\subsection{Forward modelling}\label{sec:forward-modelling}
To fit the MS to our data, we closely follow the Markov chain Monte Carlo method of \citet{2018A&A...615A.146P}. This method fits the parameters of the form of the MS being fitted as well as the scatter of the MS.

For each observed M$_{\star}$, a random SFR is drawn from a Gaussian distribution centred on the SFR of the MS being tested at that M$_{\star}$. The standard deviation of the Gaussian distribution is the scatter of the MS at that step. The Gaussian distribution is truncated such that it cannot produce SFRs larger (smaller) than the largest (smallest) observed SFR. Once the model SFRs have been generated, both the SFR and M$_{\star}$ are perturbed by adding a second random number generated from a Gaussian centred on zero and with a standard deviation of 0.25~dex. This is done to simulate the uncertainty in SFR and M$_{\star}$.

The model data is then compared to the observed data. The real and simulated data are binned into mass bins of width 0.25~dex. The means and standard deviations of the SFR in these bins are then calculated and the means and standard deviations of the model data are compared to their counterparts from the observed data. The greater the difference, the less likely the model is an accurate representation of the observed data.

In this work, we fit two forms of the MS to the data: a linear form \citep{2012ApJ...754L..29W}
\begin{equation}\label{eqn:ms-linear}
    S = \alpha\Big(\log(M_{\star}) - 10.5\Big) + \beta,
\end{equation}
where S is $\log(\mathrm{SFR}/M_{\odot}\mathrm{yr}^{-1})$, $\alpha$ is the slope and $\beta$ is the normalisation, and a form with a high-mass turn-over \citep{2015ApJ...801...80L}
\begin{equation}\label{eqn:ms-turn}
    S = S_{0} - \log\bigg(1 + \Bigg(\frac{M_{\star}}{M_{0}}\Bigg)^{-\gamma}\bigg),
\end{equation}
where $S_{0}$ is the value of S that the function approaches at high-mass, $M_{0}$ is the turn-over mass in M$_{\odot}$ and $\gamma$ is the low mass slope.

\subsection{Mass completeness}\label{sec:mass-complete}
When determining the mass completeness limit, we follow \citet{2010A&A...523A..13P} and determine this limit empirically.
We determine the mass completeness limit for each selection method and in each redshift bin before the star-forming galaxies are selected. For example, for the galaxies with NUV, r, and J band magnitudes, we determine the mass limit of the galaxies before we select the star-forming galaxies that meet the NUVrJ criteria.

Using the K$_{s}$ band limiting magnitude of 23.8 \citep{2013MNRAS.428.1281J}, the limit of observable mass for a galaxy ($M_{lim}$) can be determined per object with
\begin{equation}\label{eqn:mass_lim}
    \log{M_{lim}} = \log{M} - 0.4(K_{s,~lim} - K_{s}),
\end{equation}
where $M$ is the galaxy's mass in M$_{\odot}$, $K_{s,~lim}$ is the limiting K$_{s}$-band magnitude and $K_{s}$ is the observed K$_{s}$-band magnitude. The limiting mass of the sample is then the $M_{lim}$ that 90\% of the faintest 20\% of galaxies have $M_{lim}$ below. The mass completeness limits derived and used in this work are shown in Table \ref{tab:mass_lim}.

\begin{table*}
    \centering
    \caption{Mass completeness limits, in log(M$_{\odot}$), used in this work and derived with Eq. \ref{eqn:mass_lim}.}
    \begin{tabular}{c|ccccccc}
        \hline
        \hline
         & sSFR & NUVrJ & NUVrK & u-r & UVJ & BPT & D4000 \\
        \hline
        $0.50 \leq z < 0.62$ & 8.8 & 8.8 & 8.8 & 8.8 & 8.8 & 8.74 & 8.8 \\
        $0.62 \leq z < 0.72$ & 8.91 & 8.91 & 8.91 & 8.91 & 8.91 & 8.87 & 8.91 \\
        $0.72 \leq z < 0.85$ & 9.04 & 9.04 & 9.04 & 9.04 & 9.04 & 8.99 & 9.04 \\
        $0.85 \leq z < 1.20$ & 9.21 & 9.21 & 9.21 & 9.21 & 9.21 & 9.01 & 9.21 \\
        \hline
    \end{tabular}
    \tablefoot{Mass completeness limits are derived using the star-forming and quiescent galaxies together.}
    \label{tab:mass_lim}
\end{table*}

The VIPERS sample's mass completeness has been determined previously using i-band data in \citet{Davidzon2016} also using the \citet{2010A&A...523A..13P} method. This i-band mass completeness limit was found to be $10.18 < $log(M$_{\star}$/M$_{\odot}$)$ < 10. 86$ depending on the redshift of the galaxies and if they are star-forming or quiescent. These limits are close to, or just below the mass at which the MS has been seen to turn over \citep[e.g.][]{2015ApJ...801...80L, 2016ApJ...817..118T}. Thus, cutting at the i-band limit may hide or remove any evidence of a high-mass turn-over in the MS. \citet{2016A&A...590A.103M} also derive the K$_{s}$ band mass completeness limits, again using the \citet{2010A&A...523A..13P} method. However, \citet{2016A&A...590A.103M} uses a K$_{s}$ magnitude limit of 22 so find a mass completeness limit approximately 0.5~dex higher than the limit found in this work.

\section{Results}\label{sec:result}
\begin{figure}
	\resizebox{\hsize}{!}{\includegraphics{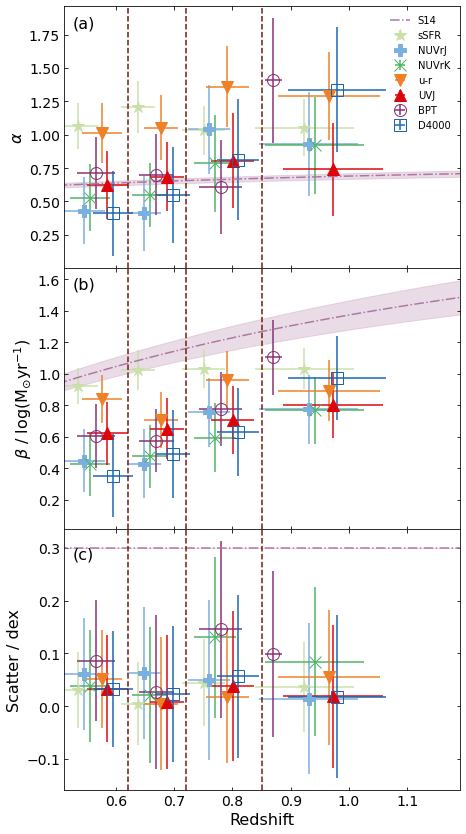}}
	\caption{Most likely (a) slope ($\alpha$), (b) normalisation ($\beta$), and (c) scatter of Eq. \ref{eqn:ms-linear} for sSFR (light green stars), NUVrJ (light blue pluses),  NVUrK (dark green crosses), u-r (orange, downward pointing triangles), UVJ (red, upward pointing triangles), BPT (empty purple circles), and D4000 (empty dark blue squares) selected galaxies. Redshifts of markers are offset for clarity. Vertical dashed lines indicate the edges of the redshift bins. Redshift and their uncertainties are average redshift of the sample and sample standard deviation. Compilation results of \citet[][S14, lilac dot-dashed line]{2014ApJS..214...15S} are included for comparison, using their intrinsic scatter.}
	\label{fig:3param}
\end{figure}

\begin{figure}
	\resizebox{0.93\hsize}{!}{\includegraphics{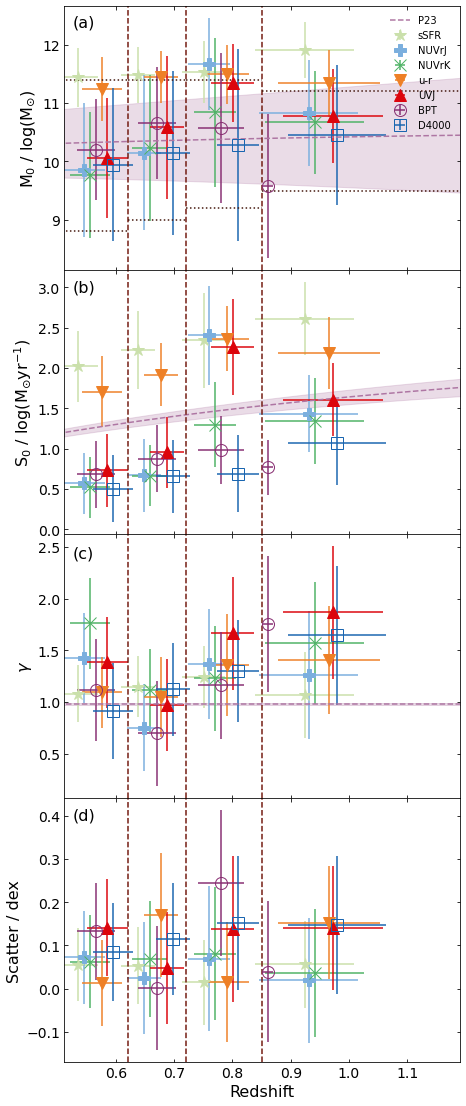}}
	\caption{Most likely (a) turn-over mass ($M_{0}$), (b) SFR that the function approaches at high-mass ($S_{0}$), (c) low mass slope ($\gamma$), and (d) scatter of Eq. \ref{eqn:ms-turn} for sSFR (light green stars), NUVrJ (light blue pluses),  NVUrK (dark green crosses), u-r (orange, downward pointing triangles), UVJ (red, upward pointing triangles), BPT (empty purple circles), and D4000 (empty dark blue squares) selected galaxies. Redshifts of markers are offset for clarity. Vertical dashed lines indicate the edges of the redshift bins. Horizontal dotted lines in (a) indicate the range within which M$_{0}$ must fall for the MS to be considered to have a turn-over. Redshift and their uncertainties are average redshift of the sample and sample standard deviation. Compilation results of \citet[][P23, lilac dashed line]{2023MNRAS.519.1526P} are included for comparison. \citet{2023MNRAS.519.1526P} do not provide a scatter in the MS so this is omitted from panel d.}
	\label{fig:4param}
\end{figure}

\begin{figure*}
	\resizebox{0.94\hsize}{!}{\includegraphics{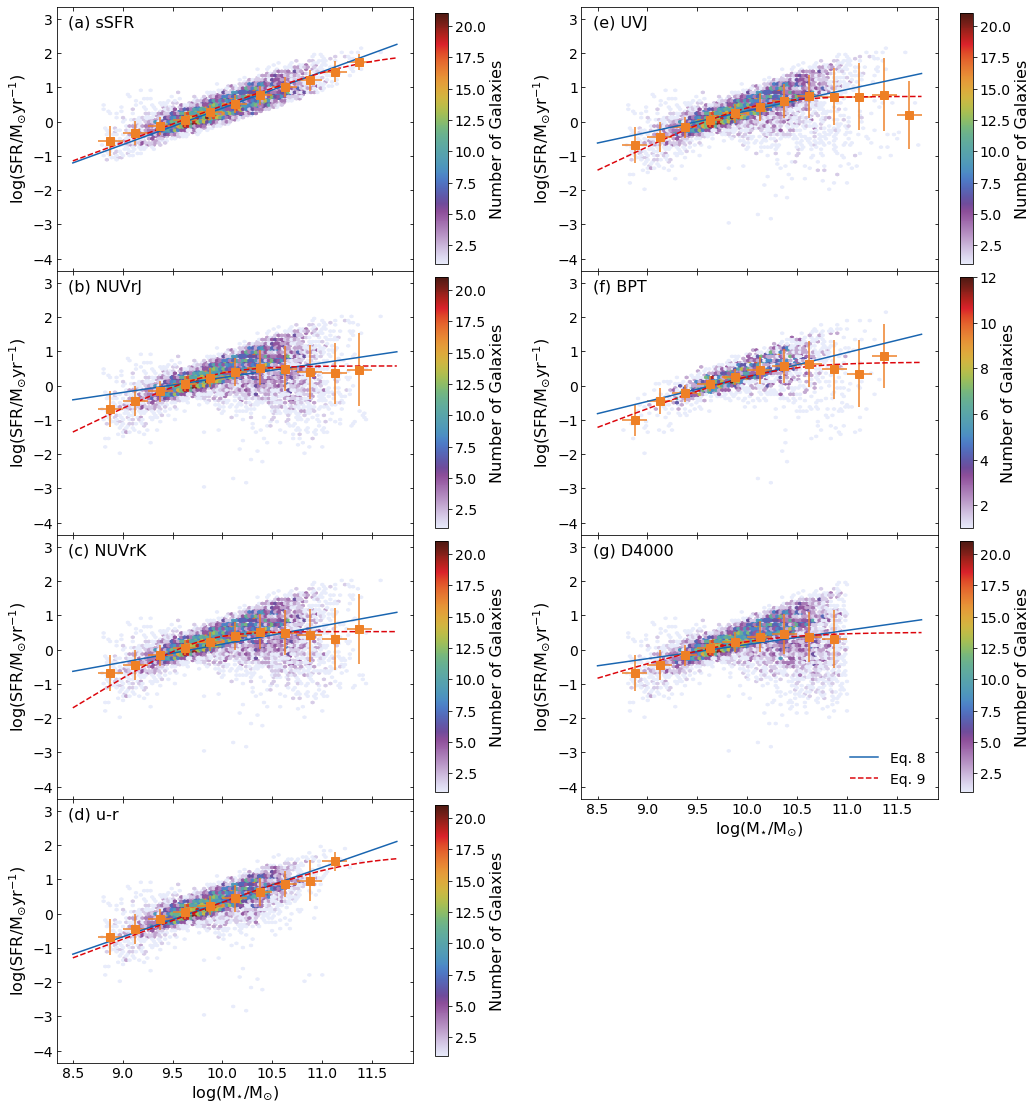}}
	\caption{Most likely linear (Eq. \ref{eqn:ms-linear}, blue line) MS and turn-over (Eq. \ref{eqn:ms-turn}, red dashed line) MS for star-forming galaxies at $0.5 \leq z < 0.62$ selected with (a) sSFR, (b) NUVrJ, (c) NUVrK, (d) u-r, (e) UVJ, (f) BPT, and (g) D4000. The mean and standard deviation of SFR within a M$_{\star}$ bin of width 0.25~dex, used during the forward modelling described in Sect. \ref{sec:forward-modelling}, are shown as orange squares and their associated vertical error bar. Horizontal error bar indicates the width of the M$_{\star}$ bin.}
	\label{fig:34paramMS}
\end{figure*}

The most likely parameters when fitting with Eq. \ref{eqn:ms-linear} are shown in Fig. \ref{fig:3param} and Table \ref{tab:3param} while the most likely parameters when fitting with Eq. \ref{eqn:ms-turn} are shown in Fig. \ref{fig:4param} and Table \ref{tab:4param}. Figure \ref{fig:34paramMS} shows the best-fit MSs in the SFR-M$_{\star}$ plane for galaxies at $0.50 \leq z < 0.62$. Similar plots for the higher redshift bins can be found in Appendix \ref{app:ms-plots}.

To determine if a MS has a turn-over, or not, we require the turn-over mass to be within the fit-able range of masses of the star-forming galaxies, and the $\chi^{2}$ for the fit with the turn-over MS (Eq. \ref{eqn:ms-turn}) to be lower than the $\chi^{2}$ of the fit with the linear MS (Eq. \ref{eqn:ms-linear}). Thus, we require $M_{0}$, the turn-over mass, to be in the ranges: $8.8 < \log(M_{0}/M_{\odot}) < 11.4$ at $0.50 \leq z < 0.62$, $9.0 < \log(M_{0}/M_{\odot}) < 11.4$ at $0.62 \leq z < 0.72$, $9.2 < \log(M_{0}/M_{\odot}) < 11.4$ at $0.72 \leq z < 0.85$, and $9.5 < \log(M_{0}/M_{\odot}) < 11.2$ at $0.85 \leq z < 1.20$. The lower limits were selected by taking the largest lowest mass of any star-forming sample (i.e. after removing the quiescent galaxies) and rounding up to 1 decimal place and the upper limits were selected by taking the smallest largest mass of any star-forming sample except D4000, as the D4000 sample is limited to log(M$_{\star}$/M$_{\odot}$) $<$ 11.0, and rounding down to 1 decimal place.

\begin{table*}
    \centering
    \caption{Most likely parameters when fitting to Eq. \ref{eqn:ms-linear}.}
    \label{tab:3param}
    \begin{tabular}{lcccccccccc}
        \hline
        \hline
        Sample & $n$ & z & $\sigma_{z}$ & $\alpha$ & $\sigma_{\alpha}$ & $\beta$ & $\sigma_{\beta}$ & Scatter & $\sigma_{Scatter}$ & $\chi^{2}$ \\
        \hline
        \multirow{4}{*}{sSFR} & 4777 & 0.57 & 0.03 & 1.06 & 0.17 & 0.92 & 0.12 & 0.03 & 0.07 & 8.65\tablefootmark{a}\\
         & 4770 & 0.67 & 0.03 & 1.21 & 0.19 & 1.02 & 0.13 & 0.00 & 0.08 & 9.79\tablefootmark{a}\\
         & 4487 & 0.78 & 0.04 & 1.03 & 0.19 & 1.03 & 0.12 & 0.04 & 0.08 & 10.38\\
         & 4026 & 0.96 & 0.09 & 1.05 & 0.21 & 1.03 & 0.13 & 0.04 & 0.09 & 9.67\\
        \hline
        \multirow{4}{*}{NUVrJ} & 6043 & 0.57 & 0.03 & 0.43 & 0.25 & 0.45 & 0.20 & 0.06 & 0.11 & 6.09\\
         & 6237 & 0.67 & 0.03 & 0.41 & 0.28 & 0.43 & 0.22 & 0.06 & 0.13 & 5.04\\
         & 5812 & 0.78 & 0.04 & 1.04 & 0.33 & 0.75 & 0.22 & 0.05 & 0.15 & 7.14\tablefootmark{a}\\
         & 4776 & 0.95 & 0.08 & 0.93 & 0.39 & 0.77 & 0.22 & 0.01 & 0.14 & 5.85\tablefootmark{a}\\
        \hline
        \multirow{4}{*}{NUVrK} & 6055 & 0.57 & 0.03 & 0.53 & 0.25 & 0.42 & 0.20 & 0.04 & 0.11 & 6.00\\
         & 6264 & 0.67 & 0.03 & 0.55 & 0.24 & 0.47 & 0.20 & 0.02 & 0.13 & 5.69\\
         & 5803 & 0.78 & 0.04 & 0.78 & 0.37 & 0.59 & 0.22 & 0.13 & 0.15 & 5.85\\
         & 4812 & 0.95 & 0.08 & 0.92 & 0.37 & 0.77 & 0.21 & 0.08 & 0.14 & 6.61\tablefootmark{a}\\
        \hline
        \multirow{4}{*}{u-r} & 4762 & 0.57 & 0.03 & 1.01 & 0.23 & 0.84 & 0.15 & 0.05 & 0.09 & 5.83\tablefootmark{a}\\
         & 4800 & 0.67 & 0.03 & 1.05 & 0.24 & 0.71 & 0.18 & 0.00 & 0.13 & 6.10\tablefootmark{a}\\
         & 4697 & 0.78 & 0.04 & 1.36 & 0.30 & 0.96 & 0.19 & 0.02 & 0.13 & 5.71\tablefootmark{a}\\
         & 4211 & 0.96 & 0.09 & 1.29 & 0.33 & 0.89 & 0.19 & 0.05 & 0.13 & 5.22\tablefootmark{a}\\
        \hline
        \multirow{4}{*}{UVJ} & 5307 & 0.57 & 0.03 & 0.62 & 0.25 & 0.62 & 0.20 & 0.03 & 0.10 & 5.75\tablefootmark{a}\\
         & 5389 & 0.67 & 0.03 & 0.69 & 0.26 & 0.65 & 0.19 & 0.01 & 0.13 & 5.75\\
         & 5194 & 0.78 & 0.04 & 0.80 & 0.35 & 0.71 & 0.22 & 0.04 & 0.14 & 5.56\tablefootmark{a}\\
         & 4481 & 0.95 & 0.09 & 0.74 & 0.35 & 0.80 & 0.21 & 0.02 & 0.14 & 6.48\\
        \hline
        \multirow{4}{*}{BPT} & 2044 & 0.57 & 0.03 & 0.71 & 0.27 & 0.61 & 0.20 & 0.09 & 0.11 & 6.55\\
         & 1058 & 0.67 & 0.03 & 0.70 & 0.30 & 0.58 & 0.20 & 0.03 & 0.15 & 2.63\tablefootmark{a}\\
         & 1371 & 0.78 & 0.04 & 0.61 & 0.35 & 0.77 & 0.24 & 0.15 & 0.17 & 4.91\\
         & 87 & 0.87 & 0.01 & 1.41 & 0.47 & 1.10 & 0.24 & 0.10 & 0.16 & 2.63\tablefootmark{a}\\
        \hline
        \multirow{4}{*}{D4000} & 5947 & 0.57 & 0.03 & 0.41 & 0.32 & 0.35 & 0.26 & 0.03 & 0.11 & 5.36\tablefootmark{a}\\
         & 6050 & 0.67 & 0.03 & 0.55 & 0.36 & 0.49 & 0.28 & 0.02 & 0.13 & 4.99\\
         & 5625 & 0.78 & 0.04 & 0.81 & 0.45 & 0.63 & 0.28 & 0.06 & 0.15 & 4.77\tablefootmark{a}\\
         & 4388 & 0.95 & 0.08 & 1.34 & 0.47 & 0.97 & 0.27 & 0.02 & 0.15 & 5.15\\
        \hline
    \end{tabular}
    \tablefoot{
            \tablefoottext{a}{$\chi^{2}$ is lower than when fitted with Eq. \ref{eqn:ms-turn}.}\\
        $n$ is the number of galaxies in the redshift bin, $\alpha$ is the slope, $\beta$ is the normalisation, $\sigma_{x}$ is the uncertainty on $x$, and $\chi^{2}$ is the $\chi^{2}$ of the best fit. Redshift and their uncertainties are average redshift of the sample and sample standard deviation.
	}
\end{table*}

\begin{table*}
    \centering
    \caption{Most likely parameters when fitting to Eq. \ref{eqn:ms-turn}.}
    \label{tab:4param}
    \begin{tabular}{lcccccccccccccc}
        \hline
        \hline
        Sample & $n$ & z & $\sigma_{z}$ & $M_{0}$ & $\sigma_{M_{0}}$ & $S_{0}$ & $\sigma_{S_{0}}$ & $\gamma$ & $\sigma_{\gamma}$ & Scatter & $\sigma_{Scatter}$ & $\chi^{2}$\\
        \hline
        \multirow{4}{*}{sSFR} & 4777 & 0.57 & 0.03 & 11.44 & 0.50 & 2.02 & 0.44 & 1.08 & 0.27 & 0.06 & 0.08 & 8.67\\
         & 4770 & 0.67 & 0.03 & 11.48 & 0.48 & 2.22 & 0.49 & 1.15 & 0.29 & 0.05 & 0.09 & 10.04\\
         & 4487 & 0.78 & 0.04 & 11.53 & 0.53 & 2.34 & 0.58 & 1.24 & 0.30 & 0.01 & 0.10 & 10.24\tablefootmark{b}\\
         & 4026 & 0.96 & 0.09 & 11.90 & 0.48 & 2.61 & 0.45 & 1.07 & 0.41 & 0.06 & 0.10 & 9.24\tablefootmark{b}\\
        \hline
        \multirow{4}{*}{NUVrJ} & 6043 & 0.57 & 0.03 & 9.85\tablefootmark{a} & 1.15 & 0.57 & 0.38 & 1.42 & 0.44 & 0.07 & 0.11 & 5.73\tablefootmark{b}\\
         & 6237 & 0.67 & 0.03 & 10.15\tablefootmark{a} & 1.33 & 0.67 & 0.45 & 0.75 & 0.42 & 0.02 & 0.13 & 4.96\tablefootmark{b}\\
         & 5812 & 0.78 & 0.04 & 11.66 & 0.79 & 2.41 & 0.62 & 1.37 & 0.53 & 0.07 & 0.17 & 7.34\\
         & 4776 & 0.95 & 0.08 & 10.83\tablefootmark{a} & 0.91 & 1.43 & 0.48 & 1.26 & 0.62 & 0.02 & 0.14 & 5.88\\
        \hline
        \multirow{4}{*}{NUVrK} & 6055 & 0.57 & 0.03 & 9.76\tablefootmark{a} & 1.08 & 0.52 & 0.38 & 1.76 & 0.44 & 0.06 & 0.11 & 5.76\tablefootmark{b}\\
         & 6264 & 0.67 & 0.03 & 10.23\tablefootmark{a} & 1.25 & 0.67 & 0.37 & 1.11 & 0.40 & 0.07 & 0.13 & 5.39\tablefootmark{b}\\
         & 5803 & 0.78 & 0.04 & 10.84\tablefootmark{a} & 1.27 & 1.30 & 0.53 & 1.23 & 0.51 & 0.08 & 0.15 & 5.80\tablefootmark{b}\\
         & 4812 & 0.95 & 0.08 & 10.67\tablefootmark{a} & 0.88 & 1.34 & 0.53 & 1.57 & 0.59 & 0.04 & 0.15 & 6.74\\
        \hline
        \multirow{4}{*}{u-r} & 4762 & 0.57 & 0.03 & 11.24\tablefootmark{a} & 0.55 & 1.71 & 0.44 & 1.09 & 0.34 & 0.01 & 0.10 & 5.85\\
         & 4800 & 0.67 & 0.03 & 11.45 & 0.44 & 1.92 & 0.39 & 1.05 & 0.39 & 0.17 & 0.14 & 6.13\\
         & 4697 & 0.78 & 0.04 & 11.49\tablefootmark{a} & 0.50 & 2.37 & 0.41 & 1.36 & 0.49 & 0.02 & 0.14 & 5.90\\
         & 4211 & 0.96 & 0.09 & 11.34 & 0.57 & 2.19 & 0.45 & 1.40 & 0.52 & 0.15 & 0.13 & 5.34\\
        \hline
        \multirow{4}{*}{UVJ} & 5307 & 0.57 & 0.03 & 10.05\tablefootmark{a} & 1.03 & 0.73 & 0.45 & 1.38 & 0.44 & 0.14 & 0.11 & 5.85\\
         & 5389 & 0.67 & 0.03 & 10.58\tablefootmark{a} & 1.22 & 0.95 & 0.44 & 0.97 & 0.44 & 0.05 & 0.13 & 4.83\tablefootmark{b}\\
         & 5194 & 0.78 & 0.04 & 11.34\tablefootmark{a} & 0.67 & 2.26 & 0.60 & 1.66 & 0.55 & 0.14 & 0.17 & 7.40\\
         & 4481 & 0.95 & 0.09 & 10.78\tablefootmark{a} & 0.80 & 1.61 & 0.45 & 1.87 & 0.64 & 0.14 & 0.14 & 5.72\tablefootmark{b}\\
        \hline
        \multirow{4}{*}{BPT} & 2044 & 0.57 & 0.03 & 10.20\tablefootmark{a} & 0.87 & 0.68 & 0.42 & 1.12 & 0.50 & 0.13 & 0.11 & 6.18\tablefootmark{b}\\
         & 1058 & 0.67 & 0.03 & 10.66\tablefootmark{a} & 0.96 & 0.88 & 0.42 & 0.70 & 0.51 & 0.00 & 0.14 & 2.77\\
         & 1371 & 0.78 & 0.04 & 10.57\tablefootmark{a} & 1.29 & 0.99 & 0.42 & 1.16 & 0.52 & 0.24 & 0.17 & 4.68\tablefootmark{b}\\
         & 87 & 0.87 & 0.01 & 9.57\tablefootmark{a} & 1.23 & 0.77 & 0.34 & 1.76 & 0.66 & 0.04 & 0.16 & 2.83\\
        \hline
        \multirow{4}{*}{D4000} & 5947 & 0.57 & 0.03 & 9.94\tablefootmark{a} & 1.31 & 0.50 & 0.42 & 0.91 & 0.46 & 0.08 & 0.11 & 5.36\\
         & 6050 & 0.67 & 0.03 & 10.14\tablefootmark{a} & 1.40 & 0.66 & 0.45 & 1.12 & 0.45 & 0.11 & 0.13 & 4.94\tablefootmark{b}\\
         & 5625 & 0.78 & 0.04 & 10.28\tablefootmark{a} & 1.64 & 0.69 & 0.48 & 1.30 & 0.49 & 0.15 & 0.15 & 4.77\\
         & 4388 & 0.95 & 0.08 & 10.45\tablefootmark{a} & 1.20 & 1.07 & 0.52 & 1.64 & 0.67 & 0.15 & 0.16 & 5.07\tablefootmark{b}\\
        \hline
    \end{tabular}
    \tablefoot{
		\tablefoottext{a}{$M_{0}$ is inside of accepted range for turn-over, see text for ranges.}
            \tablefoottext{b}{$\chi^{2}$ is lower than when fitted with Eq. \ref{eqn:ms-linear}.\\
            $n$ is the number of galaxies in the redshift bin, $S_{0}$ is the value of S that the function approaches at high-mass, $M_{0}$ is the turn-over mass in log(M$_{\odot}$), $\gamma$ is the low mass slope, $\sigma_{x}$ is the uncertainty on $x$, $\chi^{2}$ is the $\chi^{2}$ of the best fit, and TO is if there is evidence of a high-mass turn-over (Y) or not (N). Redshift and their uncertainties are average redshift of the sample and sample standard deviation.}
	}
\end{table*}

\subsection{sSFR}
Fitting Eq. \ref{eqn:ms-linear} to the sSFR selected galaxies, we find that the slope does not decrease significantly with redshift, with only a slight steepening at $0.62 \leq z < 0.72$, as can be seen in Fig. \ref{fig:3param}a (light green stars). The normalisation is seen to increase slightly between the first two bins, from $0.92 \pm 0.12$ to $1.03 \pm 0.12$, before becoming constant. The scatter remains approximately constant at $\approx0.08$~dex.

When fitting Eq. \ref{eqn:ms-turn} to the same sample, we find that $M_{0}$ rises across the entire redshift range, as shown in Fig. \ref{fig:4param}a (light green stars). The same trend is seen with $S_{0}$, with the asymptotic SFR increasing with redshift from $z = 0.50$ to $z = 1.20$. The low mass slope, $\gamma$, becomes steeper over $0.50 \leq z < 0.85$ before shallowing in the highest redshift bin, different to the trend seen when fitting Eq. \ref{eqn:ms-linear}. The scatter, however, behaves similarly to when fitting Eq. \ref{eqn:ms-linear}, remaining consistent within error at all redshift. We find $M_{0}$ does not fall within the fit-able range at any redshift and so we find no evidence of a high-mass turn-over for the sSFR sample.

\subsection{NUVrJ}
For the NUVrJ selection, we find that the slope of the linear MS is approximately constant at $z < 0.72$ before increasing into the $0.72 \leq z < 0.85$ redshift bin and again remaining approximately constant. The normalisation follows a similar trend, as shown in Fig. \ref{fig:3param}b (light blue pluses). The scatter shows a slight decrease over the entire redshift range studied but remains constant within error.

For this sample, we find that the $M_{0}$ of the turn-over MS increases with redshift from $z = 0.50$ to $z = 0.85$ before dropping out to $z = 1.20$. The only bin that has $M_{0}$ outside of the fit-able range is $0.72 \leq z < 0.85$. Of the bins with $M_{0}$ in the fit-able range, only the lower two redshift bins have a lower $\chi^{2}$ than the linear fits, and hence NUVrJ only shows evidence of a turn-over at $0.50 \leq z < 0.72$. $S_{0}$ follows the same trend as $M_{0}$: rising over $0.50 \leq z < 0.85$ before falling again. $\gamma$ generally remains approximately constant across the entire redshift range, with a lower value at $0.62 \leq z < 0.72$. For the scatter, there is a hint that it reduces slightly with redshift but it remains consistent within error, as can be seen in Fig. \ref{fig:4param}d (light blue pluses).

\subsection{NUVrK}
Despite being a very similar selection method to NUVrJ, the NUVrK selection does not follow the NUVrJ trends for the linear MS. The slope is seen to increase across the redshift range, from $0.53 \pm 0.25$ to $0.92 \pm 0.37$. The normalisation also increases over the full redshift range. The scatter of the NUVrK selection also differs from the NUVrJ selection in that there is a very slight hint of an increase with redshift but again it remains consistent within error, as seen in Fig. \ref{fig:3param}c (dark green crosses).

For the turn-over MS, the trends for NUVrK are similar to those of NUVrJ as may be expected considering the similarity of the samples. The $M_{0}$ sees an increase with redshift from $z = 0.50$ to $z = 0.85$ before dropping out to $z = 1.20$. Unlike the NUVrJ sample, however, the $M_{0}$ of the NUVrK sample remain in the fit-able range at all redshifts. While $M_{0}$ is in the fit-able range at all redshifts, the highest redshift bin has a lower $\chi^{2}$ for the linear MS and hence there is only evidence for a high-mass turn-over at $z < 0.82$. The asymptotic SFR of the NUVrK sample also increases across the entire redshift range and does not spike in the $0.72 \leq z < 0.85$ redshift bin, unlike the NUVrJ sample. The low mass slope values are consistent with the NUVrJ sample but there is an increasing value of $\gamma$ over $0.62 \leq z 1.20$ after falling between the first and second redshift bins. The scatter around the MS again remains consistent in all redshift bins.

\subsection{u-r}
The linear MS of the u-r selected star-forming galaxies shows a slight rise in slope with redshift with a significant rise between $0.62 \leq z < 0.72$ and $0.72 \leq z < 0.85$ and a slight fall into the highest redshift bin, as shown in Fig. \ref{fig:3param}a (orange, downward pointing triangles). The normalisation remains consistent across all redshift bins while the scatter around the MS shows a very slight increase over $0.62 \leq z < 1.20$, although the scatter remains consistent within error in all redshift bins.

The turn-over MS of the u-r selected star-forming galaxies shows a slight increase in $M_{0}$ over $0.52 \leq z < 0.85$ before decreasing slightly, but all $M_{0}$ values are consistent with one-another. All but the lowest redshift bin have $M_{0}$ outside of the fit-able range, but the $\chi^{2}$ in the lowest redshift bin is larger than the linear MS and, as a result, the u-r selection does not show evidence of a high-mass turn-over at any redshift. $S_{0}$ sees an increase in value out to $z = 0.85$ before falling slightly into the $0.85 \leq z < 1.20$ redshift bin. The low mass slope is consistent across all redshift bins but may be slightly lower in the lower two redshift bins when compared to the two higher bins. The scatter of the turn-over MS is approximately constant across the entire redshift range.

\subsection{UVJ}
For the UVJ selection, the slope is seen to increase with redshift along with the normalisation, but both the slope and normalisation are consistent within error in all redshift bins. As with the other star-forming galaxy selection methods, the scatter for the UVJ selection remains consistent within error in all redshift bins.

For the turn-over MS, $M_{0}$ shows an increase over $0.52 \leq z < 0.85$ before decreasing slightly. Unlike the u-r selection, the UVJ selection $M_{0}$ are all within the fit-able region. The $\chi^{2}$ in the $0.62 \leq z < 0.72$ and $0.85 \leq z < 1.20$ redshift bins are also lower than when fitting Eq. \ref{eqn:ms-linear} and so the UVJ selection shows evidence of a turn-over in these two redshift ranges. The trend seen for $S_{0}$ in Fig. \ref{fig:4param}a (red, upward pointing triangles) has a general increase across the entire redshift range but also a notable spike in value at $0.72 \leq z < 0.85$. $\gamma$ becomes shallower from $0.50 \leq z < 0.62$ to $0.62 \leq z < 0.72$ before becoming steeper out to $z = 1.20$. Once again, the scatter for the UVJ sample remains consistent in all redshift bins.

\subsection{BPT}
Due to the requirement of spectral lines in the BPT sample, the average redshift in the highest redshift bin is notably lower than the other samples, a result of the required lines only being observed out to $z = 0.9$. While fitting Eq. \ref{eqn:ms-linear}, we see the BPT sample has constant slope over $0.50 \leq z < 0.72$ before a very slight, but still consistent, drop in $0.72 \leq z < 0.85$. The highest redshift bin shows a large increase in slope to $1.41 \pm 0.47$. The normalisation also remains constant out to $z = 0.72$ but then rises throughout the rest of the redshift range, as seen in Fig. \ref{fig:3param}b (purple circles). The scatter, again, remains consistent within error across the entire redshift range.

The MS with a turn-over shows a rise in $M_{0}$ between the lowest and second lowest redshift bins before falling in the higher two redshift bins. All $M_{0}$ are in the fit-able region while the $0.50 \leq z < 0.62$ and the $0.72 \leq z < 0.85$ redshift bins have lower $\chi^{2}$, indicating that these two bins have evidence of a high-mass turn-over. For $S_{0}$, we see a rising asymptotic SFR with redshift, in Fig. \ref{fig:4param}b (purple circles), out to $z = 0.85$ before it falls into the highest redshift bin. The low mass slope, $\gamma$, decreases between the lowest two redshift bins before rising for the remainder of the redshift range. The BPT sample is the only sample where the scatter does not remain consistent for the turn-over MS. The scatter falls between the lowest and second lowest redshift bins as well as between the second highest and highest. Between the two intermediate redshift bins, the scatter is seen to increase.

\subsection{D4000}
For the D4000 sample, we find that the slope, $\alpha$ in Eq. \ref{eqn:ms-linear}, increases with redshift from $0.50 \leq z < 0.62$ to $0.85 \leq z < 1.20$. The normalisation also rises with redshift over the full redshift range, as can be seen in Fig. \ref{fig:3param}b (dark blue stars). Again we see a scatter than is consistent in all redshift bins.

Fitting with Eq. \ref{eqn:ms-turn}, we find that $M_{0}$ increase with redshift over $0.50 \leq z < 1.20$ and remains within the fit-able region. Like the UVJ sample, the $\chi^{2}$ is smaller for the turn-over MS in the $0.62 \leq z < 0.72$ and $0.85 \leq z < 1.20$ redshift bins, so these two redshift ranges show evidence of a high-mass turn-over. The value of $S_{0}$ also increases over the entire redshift range, along with the low mass slope, as shown in Fig. \ref{fig:4param}c (dark blue stars). The scatter of the turn-over MS remains consistent within error at all redshifts but there is a slight indication of a rise between $z = 0.50$ and $z = 0.85$.

\section{Discussion}\label{sec:discuss}
\subsection{Linear main sequence}
The slope of the linear MS is generally found to increase with redshift, with the exception of BPT at $z < 0.85$. This is inline with what is typically found, with many studies finding a slope that increases with redshift \citep[e.g.][]{2014ApJS..214...15S, 2018A&A...615A.146P}. While the majority of the samples following similar trends is to be expected, as all samples are drawn from the same survey catalogue, a decreasing slope with redshift for BPT was not expected. A decreasing slope with redshift has previously been seen \citep{2020MNRAS.499..948R}, however it it not commonly found. The \citet{2020MNRAS.499..948R} MS is derived with radio selected, star-forming galaxies that have had AGN removed. Their highest two redshift bins ($0.47 \leq z < 0.83$ and $0.83 \leq z < 1.20$) cover a similar range to this work. No reason is given for their decreasing slope. As the BPT sample is significantly smaller than the other samples, this decrease in slope we found for the BPT sample may be a result of this small sample size providing poor constraint on the MS.

A number of the slopes found in this work are above unity, with slopes at or below unity typically found or assumed for the MS \citep{2014ApJS..214...15S, 2023MNRAS.519.1526P}. All slopes for the sSFR and u-r samples are above unity, along with NUVrJ at $0.72 \leq z < 0.85$ and BPT and D4000 at $0.85 \leq z < 1.20$. It is not clear why these specific samples have a slope above unity. The sSFR and u-r samples both show no evidence of a high-mass turn-over at all redshifts, so the high slope is not a result of a turn-over MS being a better fit, although if this were true we would expect a lower slope, not a higher one. Similarly, NUVrJ at $0.72 \leq z < 0.85$ and BPT at $0.85 \leq z < 1.20$ do not show evidence of a high-mass turn-over. D4000 at $0.85 \leq z < 1.20$ does show evidence of a turn-over but again we would expect this to reduce the slope not increase it above unity.

Normalisations of the MSs are more in line with what would typically be expected: rising with redshift \citep[e.g.][]{2014ApJS..214...15S, 2023MNRAS.519.1526P}. The normalisations are typically found to be below those of the \citet{2014ApJS..214...15S} evolution. The exceptions are the sSFR at $z < 0.72$ and the BPT at $z \geq 0.85$, which have $\beta$ within error of the \citet{2014ApJS..214...15S} trend. Line with the slope, none of these three samples have evidence of a high-mass turn-over.

The scatters are lower than would typically be expected \citep[e.g.][]{2012ApJ...754L..29W, 2014ApJS..214...15S, 2016ApJ...820L...1K, 2016ApJ...817..118T, 2018A&A...615A.146P}, with the BPT scatter in the $0.72 \leq z < 0.85$ bin being the largest of all samples at all redshifts at $0.15 \pm 0.17$. The BPT sample also has the largest scatters of all the star-forming galaxy selections in the lowest and highest redshift bins. This larger scatter for the BPT samples may be a result of the relatively small sample size compared to the other samples. The small overall scatter of the MS for all samples is likely a result of the non-parametric SFH used. Using the same VIPERS galaxies with a parametric SFH from \citet{2016A&A...590A.103M} produces a linear MS with a much larger scatter, larger than the typically found 0.3~dex.

\subsection{Turn-over main sequence}
Not all the star-forming galaxy samples' main sequences show evidence of a high-mass turn-over. All galaxy samples used SFRs derived in the same way, here from SED fitting, and all MS were fitted in the same way (as described in Sect. \ref{sec:forward-modelling}). As only the selection criteria differ between samples, this would suggest that the selection method used to identify star-forming galaxies is a key component in the presence, or lack, of a high-mass turn-over. This supports previous works that arrived at the same conclusion \citep[e.g.][]{2015ApJ...801L..29R}. However, this does not exclude different SFR tracers or different MS tracers resulting in the presence, or lack, of a high-mass turn-over, as found by \citet{2019MNRAS.483.3213P}. Together, these suggest that the form of the MS is an artefact but it does not highlight which form is correct: if it is the high-mass turn-over or the lack of the high-mass turn-over that is artefact.

The redshifts where the different samples show little evidence of a high-mass turn-over are typically at lower redshifts, $z < 0.72$, although there are a few turn-overs found at higher redshifts. NUVrK, UVJ, BPT, and D4000 find one turn-over each at $z \geq 0.72$ while the other seven samples with high-mass turn-overs are all at $z < 0.72$. Stronger turn-overs of the MS are typically found at lower redshifts \citep{2014ApJ...795..104W, 2015ApJ...801...80L, 2016ApJ...817..118T, 2023MNRAS.519.1526P}, thus finding weaker, or no, turn-overs in all but the NUVrJ and NUVrK star-forming samples at lower redshifts was not to be expected. This does, however, agree with what is seen in our highest redshift bin, $0.85 \leq z < 1.20$, which only has turn-overs for UVJ and D4000.

Selecting on NUVrJ and NUVrK see turn-overs at $z < 0.72$ and $z < 0.85$, respectively. As these two samples are similar types of cuts, it is not surprising to find similar results. As the turn-over disappears at higher redshifts, these two samples support the commonly found observation that higher redshifts samples have a weaker, or no, turn-over than their lower redshift counterparts. This would suggest that the higher mass galaxies that are considered to be star-forming by these two selections are less efficient at forming stars in the older universe than in the younger universe. These two samples may also be contaminated by green valley galaxies, which may also be helping to create the high-mass turn-overs seen.

The sSFR and u-r samples are the only selections that do not contain a redshift bin that shows evidence of a turn-over, and sSFR is the only selection that does not have a redshift bin with $M_{0}$ within the fit-able region. The sSFR selection is very strict in rejecting passive galaxies. With the sSFR cut being so close the the MS and the tight scatter, it is not surprising that this sample was found to be linear. Similarly, the u-r sample does not show many low star-formation objects in the sample, with only a low number at higher masses, suggesting that the u-r selection is relatively strict in removing passive galaxies resulting in a linear MS. The u-r selection may also be removing dusty star-forming galaxies due to its sensitivity to dust attenuation. As dusty galaxies are typically high mass, this may also be contributing to the lack of a high-mass turn-over for the u-r sample.

For the UVJ, BPT, and D4000 samples, there is no redshift trend with where they see a turn-over. UVJ and D4000 see turn-overs at $0.62 \leq z < 0.72$ and $0.85 \leq z < 1.20$ while BPT sees turn-overs at $0.50 \leq z < 0.62$ and $0.72 \leq z < 0.85$. If a turn-over is a result of quiescent contamination of the star-forming sample and the BPT sample is the most conservative in selecting star-forming galaxies, it would be expected that all star-forming samples should have turn-overs at least at the same redshifts. This is evidently not the case, suggesting that the BPT sample may not necessarily be the purest star-forming sample. As the D4000 samples are known to be contaminated by quiescent galaxies \citep{2008A&A...487...89V} and our D4000 sample turn-overs also have no relation to redshift, the BPT sample may contain contaminants. Alternatively, the turn-over is not the result of quiescent contamination in the star-forming sample.

\subsection{Comparison with other works}
Here we compare our work to literature results gained using the same, or similar, star-forming galaxy selection.

The NUVrJ selected MSs of \citet{2015ApJ...801...80L, 2017A&A...605A..70D, 2020ApJ...899...58L} all show high-mass turn-overs at all redshifts. \citet{2015ApJ...801...80L} study galaxies with $z \leq 1.3$ from the multi-wavelength Cosmic Evolution Survey field \citep[COSMOS][]{2007ApJS..172....1S} and derive M$_{\star}$ and SFR with SED fitting. \citet{2017A&A...605A..70D} use $z < 6.0$ COSMOS galaxies from the \citet{2016ApJS..224...24L} catalogue and determine M$_{\star}$ and SFR by SED fitting with Le Phare \citep{1999MNRAS.310..540A, 2006A&A...457..841I}. The last of these three studies, \citet{2020ApJ...899...58L}, also uses the \citet{2016ApJS..224...24L} catalogue, combined with VLA-COSMOS 3 GHz Large Project \citep{2017A&A...602A...2S} radio data, for galaxies out to $z \approx 5.0$ and determine SFR through the radio-infrared correlation \citep{2018MNRAS.475..827M} and M$_{\star}$ through SED fitting. All three studies show an increase in normalisation and low mass slope with redshift. For the NUVrJ selected galaxies in this work, we do not find a turn-overs in our highest two redshift bins ($0.72 \leq z < 1.20$). The normalisation that we find, $\beta$ and $S_{0}$, does increase with redshift, although our low mass slope also remains approximately constant with redshift. Thus, this work is in partial agreement with the existing literature, more so at low redshift than high.

For NUVrK, \citet{2015A&A...579A...2I} find a turn-over at all redshifts ($0.2 < z < 1.4$), unlike this work that only finds turn-overs at lower redshifts ($0.50 \leq z < 0.85$). \citet{2015A&A...579A...2I} also find an increase in normalisation and low mass slope. As has been stated already, in this work we find an increase in normalisation with redshift as well as an increase in the low mass slope with redshift above $z = 0.62$. Thus, our NUVrK results are in reasonable agreement with the existing literature. The \citet{2015A&A...579A...2I} results are based on a $24~\mu m$ selected sample with M$_{\star}$s estimated by SED fitting with Le Phare while the SFR is derived from the total infrared (IR) emission summed with the total ultra-violet (UV) emission \citep{2013A&A...558A..67A}.

With UVJ selection of NEWFIRM Medium-Band Survey \citep{2011ApJ...735...86W} galaxies, \citet{2012ApJ...754L..29W} found a weak turn-over at $0.50 < z < 1.0$ that becomes more apparent as redshift increases, out to $z = 2.5$. The M$_{\star}$s in \citet{2012ApJ...754L..29W} are estimated by SED fitting with FAST \citep{2009ApJ...700..221K} and SFR is estimated by summing the total IR and UV emissions \citep{1998ARA&A..36..189K}. This turn-over evolution is not what is seen in the work, with no turn-over present at $0.50 \leq z < 0.62$ or $0.72 \leq z < 0.85$. However, as the redshift bins that we find to have high-mass turn-over are in and partially in the range $0.5 < z < 1.0$, it is possible that if we fitted to such a redshift range a turn-over would be come apparent. As a turn-over is seen in our redshift range $0.85 \leq z < 1.20$, it is also possible that a turn-over would be seen using the \citet{2012ApJ...754L..29W} $1.0 < z < 1.5$ redshift binning.

In a more recent work with UVJ selection, \citet{2014ApJ...795..104W} find a weaker turn-over as redshift increases and a non-evolving low mass slope. Our UVJ results do not have a clear redshift evolution of the turn-over MS. The $\gamma$ evolution we find also sees the low mass slope increasing with redshift from $z = 0.62$. \citet{2014ApJ...795..104W} again used FAST to fit SEDs and estimate the M$_{\star}$ while the SFR is once again determined by combining the UV and IR emission \citep{2014ApJ...795..104W} from CANDELS \citep{2011ApJS..197...35G}, 3D-HST \citep{2012ApJS..200...13B} and $24~\mu m$ observations over $0.5 < z < 2.5$.

The \citet{2018A&A...615A.146P} UVJ selected MS shows no high-mass turn-over at any redshift ($0.2 < z < 6.0$). These galaxies are far-IR selected with M$_{\star}$s and SFRs estimated by SED fitting with CIGALE \citep{2019A&A...622A.103B} with a parametric SFH. This lack of turn-over agrees with what we see for half of our redshift bins. \citet{2018A&A...615A.146P} find a slope that increases with redshift, in agreement with our results at $z > 0.62$. The inclusion of far-infrared data for the SFR estimation in \citet{2018A&A...615A.146P} may be the cause of this discrepancy. Higher redshift galaxies are more dusty \citep[e.g.][]{2011MNRAS.417.1510D, 2020A&A...644A.144D} and this may cause an under estimation of the SFR of high-mass galaxies, resulting in a shallower slope at higher redshift. 

\subsection{Form of the main sequence}
Evidently, different star-forming galaxy selections present different forms of the MS. As a result, the high-mass turn-over seen in some studies is likely a result of the method used to select the star-forming galaxies. It is likely that stricter selections results in a more linear MS, while a selection with greater chance of quiescent contamination will result in a MS with a turn-over.

\section{Summary and Conclusions}\label{sec:conclusion}
In this work we studied the effect of galaxy selection on the MS. We selected star-forming galaxies photometrically using sSFR, NUVrJ, NUVrK, u-r and UVJ, and spectroscopically using BPT and D4000 cuts from the VIPERS sample over $0.5 \leq z < 1.2$. The selected star-forming galaxies were then fitted with both a linear MS and a MS with a high-mass turn-over.

The slope of all star-forming samples were found to either remain constant or increase with redshift, the normalisation was also constant or increasing with redshift, while the scatters remained approximately constant. No galaxy samples had turn-overs at all redshifts and only the NUVrJ and NUVrK galaxy samples had a clear redshift evolution of the presence of a high-mass turn-over. The sSFR and u-r selected galaxies show no high-mass turn-over at any redshift. There is not a redshift bin that all the remaining samples all showed a high-mass turn-over.

As a result, it is apparent that the presence of a high-mass turn-over, or lack there of, is at least partially a result of the galaxy selection method. As we used the same parent sample of galaxies, if this were not the case all different selection methods used would result in either the presence or lack of a high-mass turn-over. However, we cannot exclude the possibility that there are other influences on the presence of a high-mass turn-over in the MS, such as the tracer used to fit the MS \citep[mean, mode or median][]{2019MNRAS.483.3213P} or the SFR tracer used.

\begin{acknowledgements}
    We would like to thank the referee for their thorough and thoughtful comments that helped improve the quality and clarity of this paper.
    We would like to thank L. Cielsa for providing the sfrNlevels module for CIGALE.
    W.J.P. has been supported by the Polish National Science Center project UMO-2020/37/B/ST9/00466 and by the Foundation for Polish Science (FNP).
    M.F. has been supported by the First TEAM grant of the Foundation for Polish Science No. POIR.04.04.00-00-5D21/18-00 (PI: A. Karska)
    K.M has been supported by the Polish National Science Center project UMO-2018/30/E/ST9/00082.
    This work has been partially supported by the Polish National Science Center project UMO-2018/30/M/ST9/00757.
    This paper uses data from the VIMOS Public Extragalactic Redshift Survey (VIPERS). VIPERS has been performed using the ESO Very Large Telescope, under the "Large Programme" 182.A-0886. The participating institutions and funding agencies are listed at http://vipers.inaf.it
\end{acknowledgements}

\bibliographystyle{aa} 
\bibliography{VIPERS-MS}

\begin{appendix}
\section{CIGALE model}\label{app:cigale}
For the CIGALE models, we use a non-parametric SFH \citep[sfhNlevels]{2023A&A...672A.191C}, \citet[bc03]{2003MNRAS.344.1000B} stellar population, \citet{2003PASP..115..763C} initial mass function, \citet[dustatt\_2powerlaws]{2000ApJ...539..718C} dust attenuation, \citet[dl2014]{2014ApJ...780..172D} dust emission, and \citet[fritz2006]{2006MNRAS.366..767F} AGN emission. The parameters used in CIGALE can be found in Table \ref{tab:parameters}.

\begin{table*}\label{tab:parameters}
        \caption{Parameters used for the CIGALE model SEDs used to estimate M$_{\star}$ and SFR, and the u-r, U-V, and V-J rest-frame colours.}
        \centering
        \begin{tabular}{l c c}
                \hline
                \hline
                Parameter & Value & Description\\
                \hline
                \hline
                \multicolumn{3}{c}{Star-formation history (sfhNlevels)} \\
                \hline
                & & \\
                $age$ & 500, 750, 1000, 1250, 1500, 1750, 2000, & Age of the oldest stars\\
                 & 2250, 2500, 2750, 3000, 3250, 3500, 3750, &\\
                 & 4000, 4250, 4500, 4750, 5000, 5250, 5500, &\\
                 & 5750, 6000, 6250, 6500, 6750, 7000, 7250, &\\
                 & 7500, 7750, 8000, 8250, 8500, 8750, 9000 &\\
                $age\_{1stbin}$ & 30 & Age of the 1st SFH bin \\
                $N_{bins}$ & 7 & Number of bins in the SFH\\
                $N_{SFH}$ & 10, 100, 1000, 10000, 100000, & Seeds of the models\\
                 & 1000000, 10000000 &\\
                $t_{gradent}$ & 36 & Time over which the SFH gradient is calculated\\
                $sfr_{A}$ & 1.0 & Factor controlling the amplitude of SFR\\
                normalise & True & Normalise the SFH to produce 1~M$_{\odot}$\\
                 & &\\
                \hline
                \hline
                \multicolumn{3}{c}{Stellar Emission (bc03)}\\
                \hline
                & & \\
                IMF & \citet{2003PASP..115..763C} & Initial Mass Function\\
                $Z$ & 0.02 & Metallicity (0.02 is Solar)\\
                Separation Age & 0.01 & Separation between young and old stellar populations\\
                & & \\
                \hline
                \hline
                \multicolumn{3}{c}{Dust attenuation (dustatt\_2powerlaws)}\\
                \hline
                & & \\
                A$_V^{BC}$ & 0.3, 1.2, 2.3, 3.3, 3.8 & V-band attenuation of the birth clouds\\
                Slope$_{BC}$ & -0.7 & Birth cloud attenuation power law slope\\
                BC to ISM Factor & 0.3, 0.5, 0.8, 1.0 & Ratio of the birth cloud attenuation to ISM attenuation\\
                Slope$_{ISM}$ & -0.7 & ISM attenuation power law slope\\
                & & \\
                \hline
                \hline
                \multicolumn{3}{c}{Dust emission (dl2014)}\\
                \hline
                & & \\
                $q_{PAH}$ & 0.47, 1.12, 2.50, 3.9 & Mass fraction of PAH \\
                $U_{min}$ & 5.0, 10.0, 25.0 & Minimum scaling factor of the radiation field intensity\\
                $\alpha$ & 2.0 & Dust power law slope\\
                $\gamma$ & 0.02 & Illuminated fraction\\
                & & \\
                \hline
                \hline
                \multicolumn{3}{c}{AGN template (fitz2006)}\\
                \hline
                & & \\
                $r_{ratio}$ & 60.0 & Ratio of maximum to minimum radii\\
                $\tau$ & 1.0, 6.0 & Optical depth at 9.7~$\mu$m\\
                $\beta$ & -0.5 & $\beta$ coefficient for the gas density function of the torus\tablefootmark{a}\\
                $\gamma$ & 0.0 & $\gamma$ coefficient for the gas density function of the torus\\
                Opening Angle & 100.0\degr & Opening angle of the torus\\
                $\psi$ & 0.001\degr, 89.990\degr & Angle between equatorial axis and line of sight\\
                disk type & \citet{2000ApJ...533..682C} & Disk spectrum\\
                $\delta$ & -0.36 & Power-law of $\delta$\\
                $frac_{AGN}$ & 0.0, 0.1, 0.3, 0.5, 0.7 & AGN fraction\\
                $\lambda_{fracAGN}$ & 0/0 & Wavelength range to compute the AGN fraction\\
                 & &( 0/0 is total dust luminosity)\\
                law & Small Magellanic Cloud & Extinction law of the polar dust\\
                E(B-V) & 0.03 & E(B-V) extinction in the moral direction in mag\\
                Temperature & 100K & Polar dust temperature\\
                Emissivity & 1.6 & emissivity of the polar dust\\
                & & \\
                \hline
        \end{tabular}
        \tablefoot{All ages and times are in Myr.}
\end{table*}

\section{Star-forming galaxy selection plots}\label{app:sf-plots}
\begin{figure}
    \resizebox{\hsize}{!}{\includegraphics{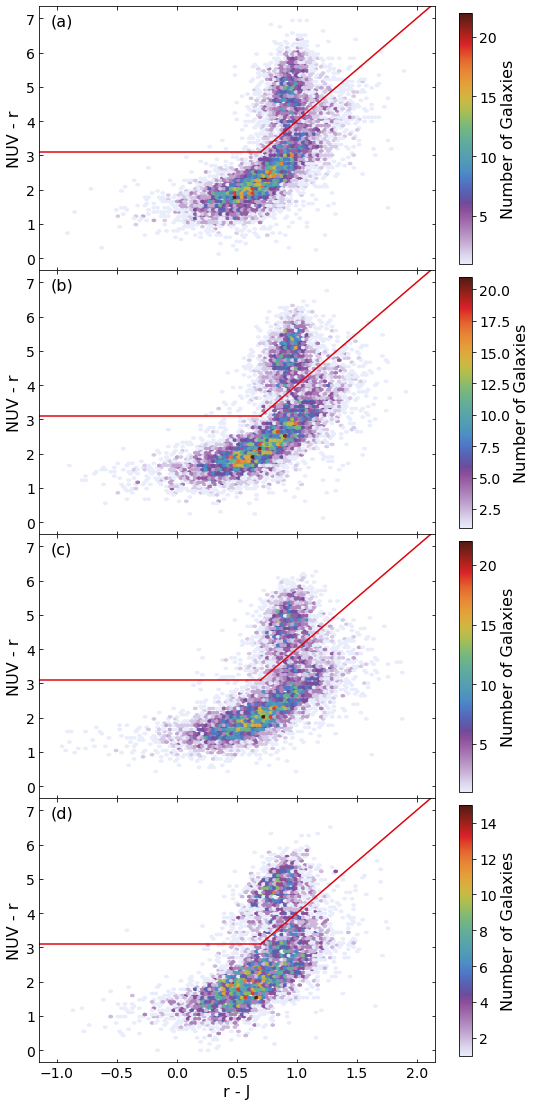}}
    \caption{Density plot of NUV-r vs r-J for galaxies with (a) $0.50 \leq z < 0.62$, (b) $0.62 \leq z < 0.72$, (c) $0.72 \leq z < 0.85$, and (d) $0.85 \leq z < 1.20$ from low density (light purple) to high density (dark red). The NUVrJ cuts are shown as a red lines.}
    \label{fig:NUVrJ-cut}
\end{figure}

\begin{figure}
    \resizebox{\hsize}{!}{\includegraphics{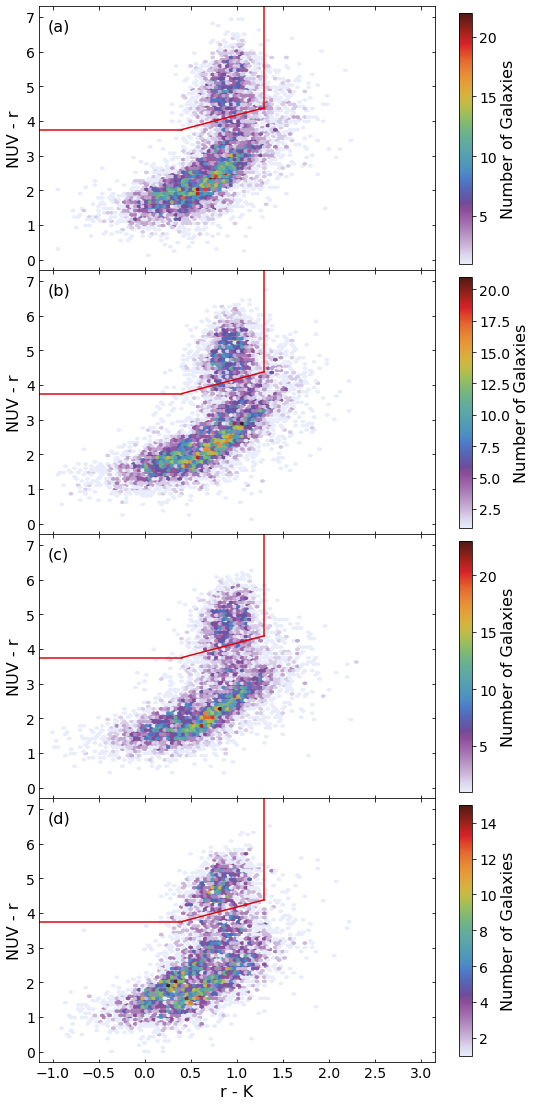}}
    \caption{Density plot of NUV-r vs r-K for galaxies with (a) $0.50 \leq z < 0.62$, (b) $0.62 \leq z < 0.72$, (c) $0.72 \leq z < 0.85$, and (d) $0.85 \leq z < 1.20$ from low density (light purple) to high density (dark red). The NUVrK cuts are shown as a red lines.}
    \label{fig:NUVrK-cut}
\end{figure}

\begin{figure}
    \resizebox{\hsize}{!}{\includegraphics{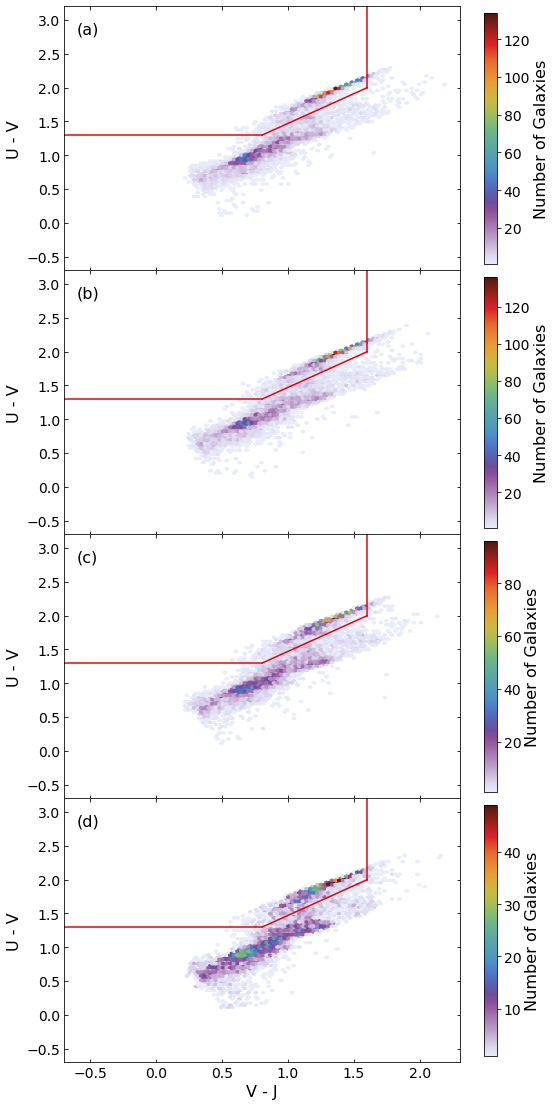}}
    \caption{Density plot of U-V vs V-J for galaxies with (a) $0.50 \leq z < 0.62$, (b) $0.62 \leq z < 0.72$, (c) $0.72 \leq z < 0.85$, and (d) $0.85 \leq z < 1.20$ from low density (light purple) to high density (dark red). The UVJ cuts are shown as a red lines.}
    \label{fig:UVJ-cut}
\end{figure}

\begin{figure}
    \resizebox{\hsize}{!}{\includegraphics{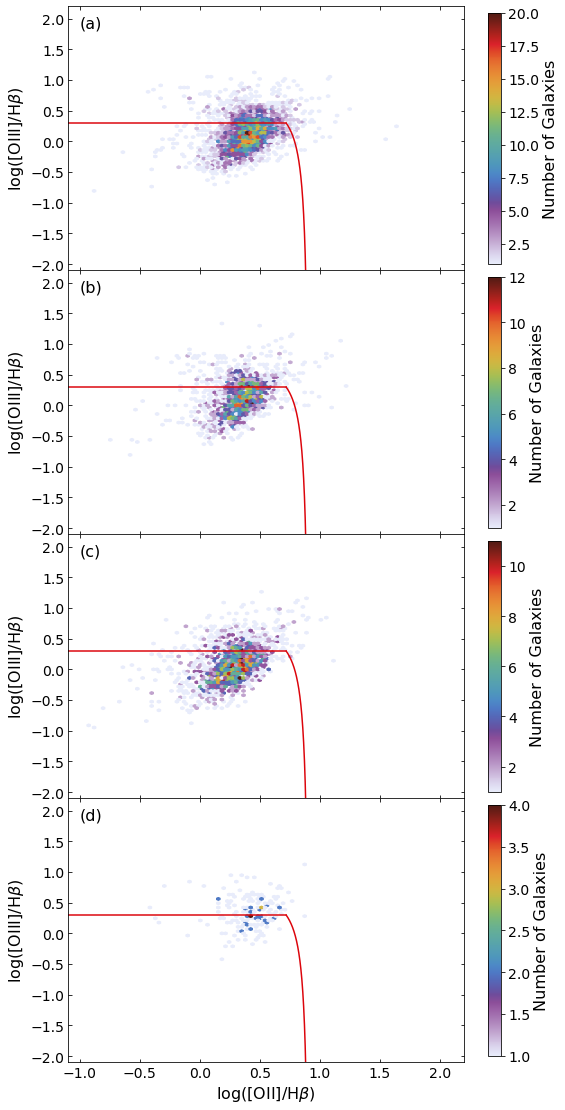}}
    \caption{Density plot of $\log([\mathrm{\ion{O}{iii}}]/\mathrm{H}\beta)$ vs $\log([\mathrm{\ion{O}{ii}}]/\mathrm{H}\beta$ for galaxies with (a) $0.50 \leq z < 0.62$, (b) $0.62 \leq z < 0.72$, (c) $0.72 \leq z < 0.85$, and (d) $0.85 \leq z < 1.20$ from low density (light purple) to high density (dark red). The BPT cuts are shown as a red lines.}
    \label{fig:BPT-cut}
\end{figure}

\begin{figure}
    \resizebox{\hsize}{!}{\includegraphics{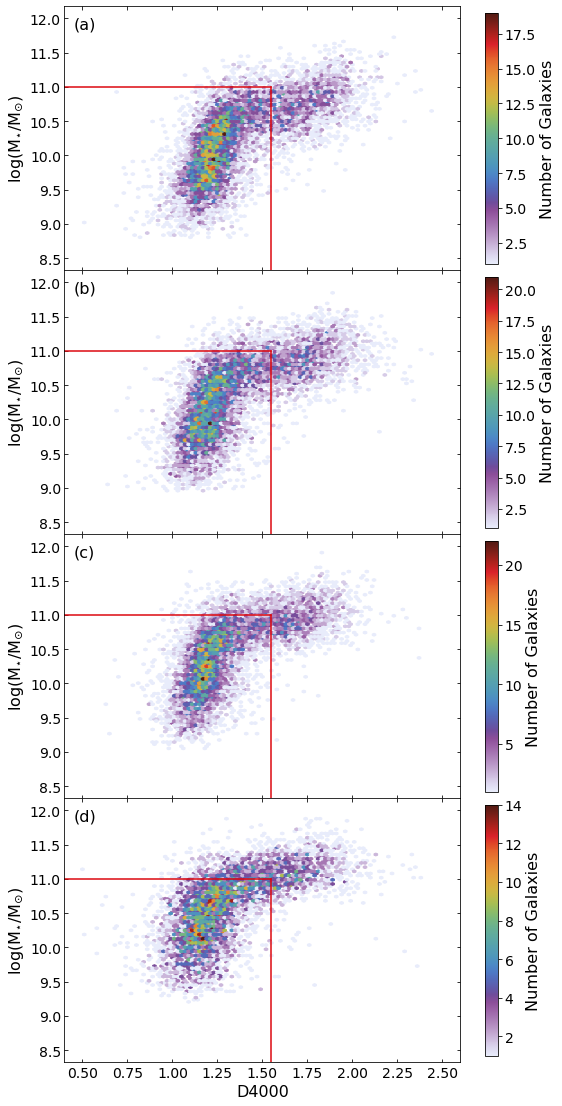}}
    \caption{Density plot of M$_{\star}$ vs D4000 for galaxies with (a) $0.50 \leq z < 0.62$, (b) $0.62 \leq z < 0.72$, (c) $0.72 \leq z < 0.85$, and (d) $0.85 \leq z < 1.20$ from low density (light purple) to high density (dark red). The D4000 cuts are shown as a red lines.}
    \label{fig:D4000-cut}
\end{figure}

Here we present the star-forming galaxy selection plots for the NUVrJ selection in Fig. \ref{fig:NUVrJ-cut}, the NUVrK selection in Fig. \ref{fig:NUVrK-cut}, the UVJ selection in Fig. \ref{fig:UVJ-cut}, the BPT selection in Fig. \ref{fig:BPT-cut}, and the D4000 selection in Fig. \ref{fig:D4000-cut}.

\section{Further main sequence plots}\label{app:ms-plots}
\begin{figure*}
	\resizebox{0.94\hsize}{!}{\includegraphics{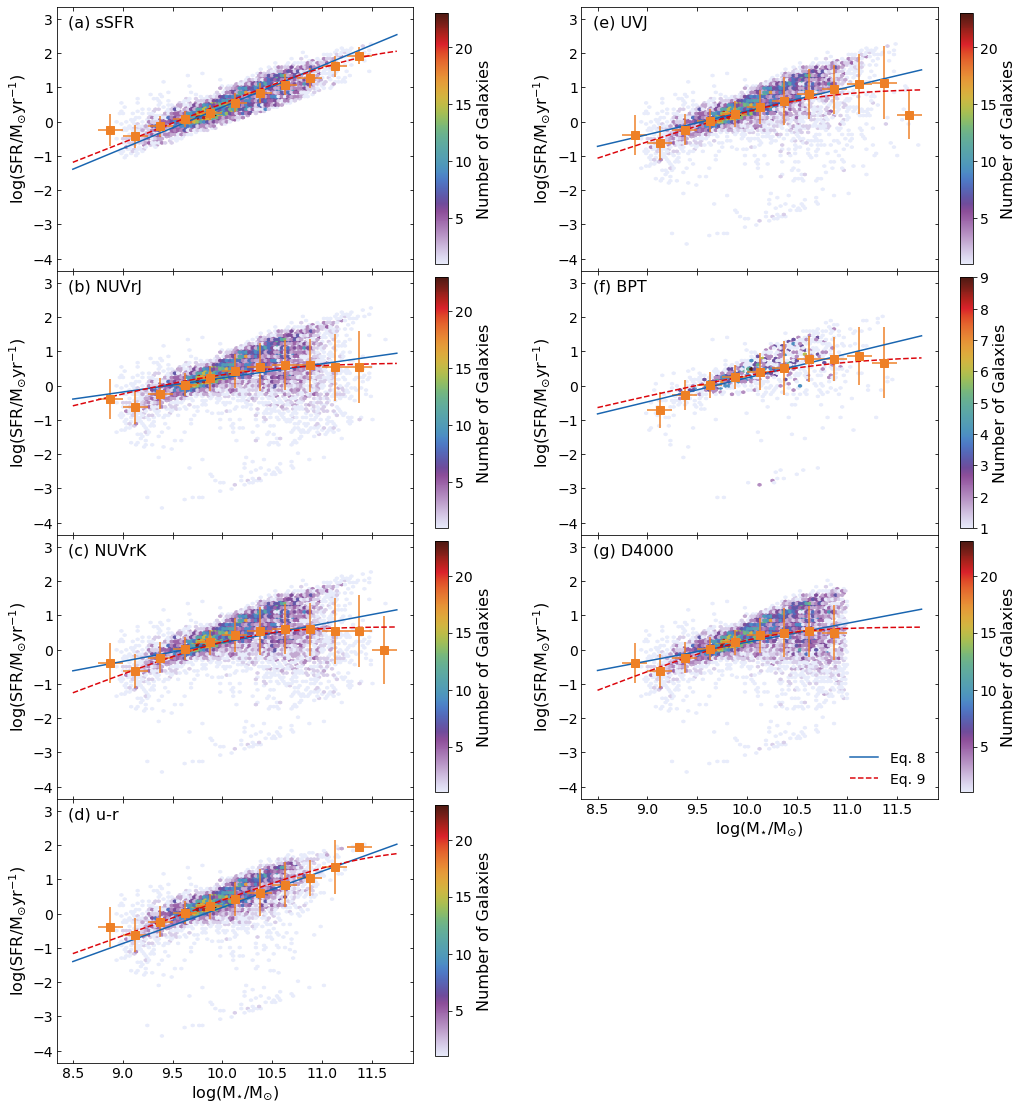}}
	\caption{Most likely linear (Eq. \ref{eqn:ms-linear}, blue line) MS and turn-over (Eq. \ref{eqn:ms-turn}, red dashed line) MS for star-forming galaxies at $0.62 \leq z < 0.72$ selected with (a) sSFR, (b) NUVrJ, (c) NUVrK, (d) u-r, (e) UVJ, (f) BPT, and (g) D4000. The mean and standard deviation of SFR within a M$_{\star}$ bin of width 0.25~dex, used during the forward modelling described in Sect. \ref{sec:forward-modelling}, are shown as orange squares and their associated vertical error bar. Horizontal error bar indicates the width of the M$_{\star}$ bin.}
	\label{fig:34paramMS-1}
\end{figure*}

\begin{figure*}
	\resizebox{0.94\hsize}{!}{\includegraphics{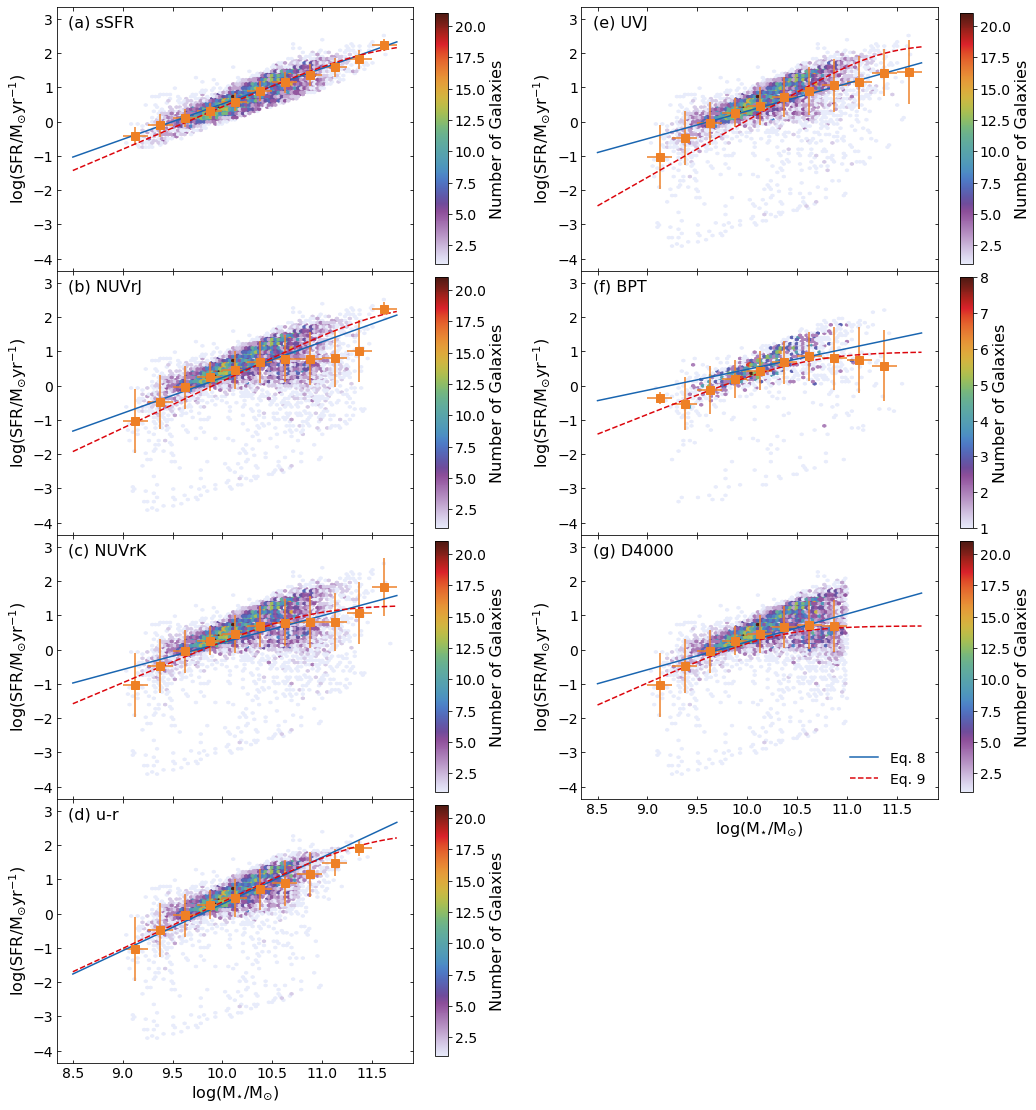}}
	\caption{As Fig. \ref{fig:34paramMS-1} but for galaxies at $0.72 \leq z < 0.85$.}
	\label{fig:34paramMS-2}
\end{figure*}

\begin{figure*}
	\resizebox{0.94\hsize}{!}{\includegraphics{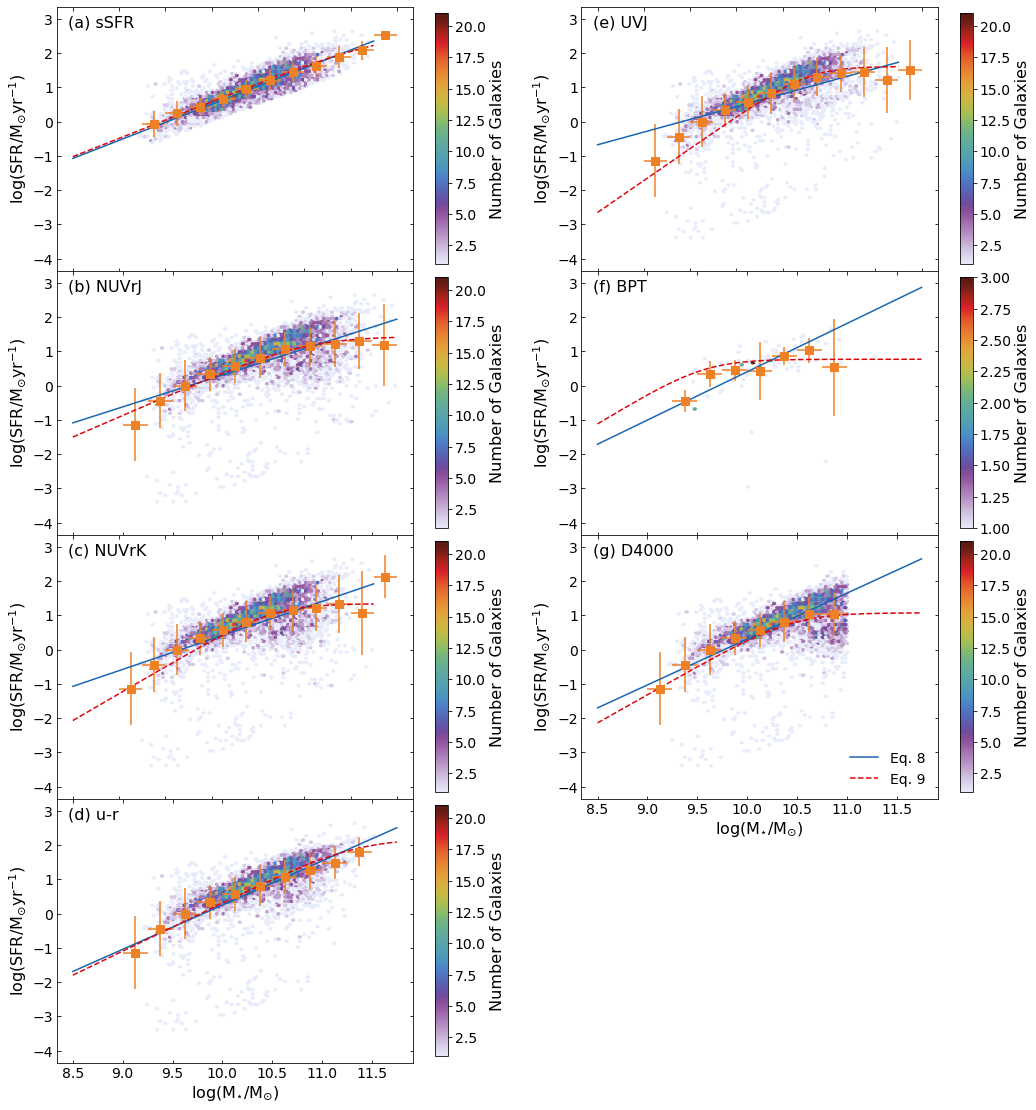}}
	\caption{As Fig. \ref{fig:34paramMS-1} but for galaxies at $0.85 \leq z < 1.20$.}
	\label{fig:34paramMS-3}
\end{figure*}

Here we present the best fit linear and turn-over MSs in the SFR-M$_{\star}$ plane for galaxies at $0.62 \leq z < 0.72$ (Fig. \ref{fig:34paramMS-1}), $0.72 \leq z < 0.85$ (Fig. \ref{fig:34paramMS-2}), and $0.85 \leq z < 1.20$ (Fig. \ref{fig:34paramMS-3}). A similar plot for galaxies at $0.50 \leq z < 0.62$ can be found in Fig. \ref{fig:34paramMS}.

\end{appendix}

\end{document}